# AI in Finance: Challenges, Techniques and Opportunities

LONGBING CAO, University of Technology Sydney, Australia

AI in finance broadly refers to the applications of AI techniques in financial businesses. This area has attracted attention for decades with both classic and modern AI techniques applied to increasingly broader areas of finance, economy and society. In contrast to either discussing the problems, aspects and opportunities of finance that have benefited from specific AI techniques and in particular some new-generation AI and data science (AIDS) areas or reviewing the progress of applying specific techniques to resolving certain financial problems, this review offers a comprehensive and dense roadmap of the overwhelming challenges, techniques and opportunities of AI research in finance over the past decades. The landscapes and challenges of financial businesses and data are firstly outlined, followed by a comprehensive categorization and a dense overview of the decades of AI research in finance. We then structure and illustrate the data-driven analytics and learning of financial businesses and data. A comparison, criticism and discussion of classic vs. modern AI techniques for finance follows. Finally, the open issues and opportunities to address future AI-empowered finance and finance-motivated AI research are discussed.

## 1 INTRODUCTION

Artificial intelligence (AI) in finance has been a research area of great interest for many decades. Classic AI-empowered finance and economics such as for traditional financial markets, trading, banking, insurance, risk, regulation and marketing has evolved to the new-generation FinTech (or Fintech, i.e., financial technology in full) enabling smart digital currencies, lending, payment, asset and wealth management, risk and regulation management, and accounting and auditing [13, 14, 32, 33, 58, 119, 136]. Here, *finance* refers to broad areas including capital markets, trading, banking, insurance, leading/loan, investment, asset/wealth management, risk management, marketing, compliance and regulation, payment, contracting, auditing, accounting, financial infrastructure, blockchain, financial operations, financial services, financial security, and financial ethics. In addition, economics and finance (EcoFin for short, which also refers to the EcoFin aspects below) are increasingly synergized by each other and with the broad AI family.

The recent EcoFin transformation and paradigm shift has been mainly driven by the new-generation AI and data science (AIDS) advancement [14, 30, 84, 149, 155], which are innovating, transforming and synthesizing financial services, economy, technology, media, communication and society [33]. AIDS consist of (1) *classic techniques* including logic, planning, knowledge representation, statistical modeling, mathematical modeling, optimization, autonomous systems, multiagent systems, expert systems (ES), decision support system (DSS), simulation, complexity science, pattern recognition, image processing, and natural language processing (NLP); and (2) *modern techniques* such as recent advances in representation learning, machine learning, optimization, data analytics, data mining and knowledge discovery, computational intelligence, event analysis, behavior informatics, social media/network analysis, and more recently deep learning, cognitive computing and quantum computing. As discussed in [33], AIDS largely defines the objectives, products and services of the new era of EcoFin and FinTech and nurtures waves of EcoFin transformations toward increasingly proactive, personalized, intelligent, interconnected, secure and trustful products and services, forming *smart FinTech* (as shown in Fig. 2 in [33]).

Fig. 1 connects the major AIDS techniques (the upper fin), EcoFin businesses (the lower fin), and their synergy (as shown in the ridge). The symbiont (the FinTech fish) fosters various synthetic areas in the 'smart FinTech' family: (1) intelligentizing the core EcoFin businesses, such as smart banking, smart insurance, smart lending, smart trading, smart wealth, smart blockchain, smart payment, and smart marketing; (2) intelligentizing their operations, services and decision-making, such as smart regulation, smart risk, smart security, smart accounting, smart auditing, smart governance, smart operations, and smart management; and (3) enabling smarter FinTech futures, such as creating even smarter design, planning, and innovations. Respective AIDS techniques are developed to enable and automate the 'smartness' of these areas, further producing the corresponding FinTech technical spectrum: BankingTech, LendTech, WealthTech, TradeTech, PayTech, InsurTech, RiskTech, and RegTech, etc. These form the concept and family of 'smart FinTech'. Interested readers can refer to [32, 33] and other related references for an introduction to these ever-evolving areas and techniques. The AIDS techniques reviewed in this paper directly enable the translation from the above smart FinTech businesses to their corresponding smart FinTech techniques.

Many review references in the literature are more or less related to AI in finance, as illustrated by the 35 or so papers cited in this paper. The existing reviews focus on the applications of (1) a specific technique or method, such as time series analysis, text mining, natural language processing, data mining, classic machine learning, evolutionary computing, computational intelligence, quantum computing, or deep learning and (2) specific business problems, such as market trend forecasting, stock price prediction, credit scoring, fraud detection, financial report analysis, pricing and hedging, marketing, consumer behavior analysis, algorithmic trading, social commerce, and Internet finance. To the best of our knowledge, no comprehensive reviews address the entire ecosystems of techniques and businesses and their synergies, an ambitious but challenging task. In [33], a comprehensive review summarizes the decades of AI research in finance from the financial application perspectives, which appears to be the first comprehensive survey on structuring, analyzing and commenting on the application ecosystem. This paper complements this review but takes a technical perspective to further summarize, structure, analyze, compare and comment on the fundamental business and data challenges in finance to AIDS, the AIDS techniques for a data-driven understanding and resolution of financial problems, and the technical gaps and opportunities for future AI research in finance. It appears to be the first attempt to draw a comprehensive but highly dense overview of the technical ecosystem of AI in finance.

## 2  AI-EMPOWERED FINANCIAL BUSINESSES AND CHALLENGES

As seen in numerous studies (e.g., [33]), AIDS techniques have been intensively applied to address widespread EcoFin and FinTech business problems and opportunities. Here, we briefly summarize EcoFin businesses and their challenges that can better benefit from AIDS.

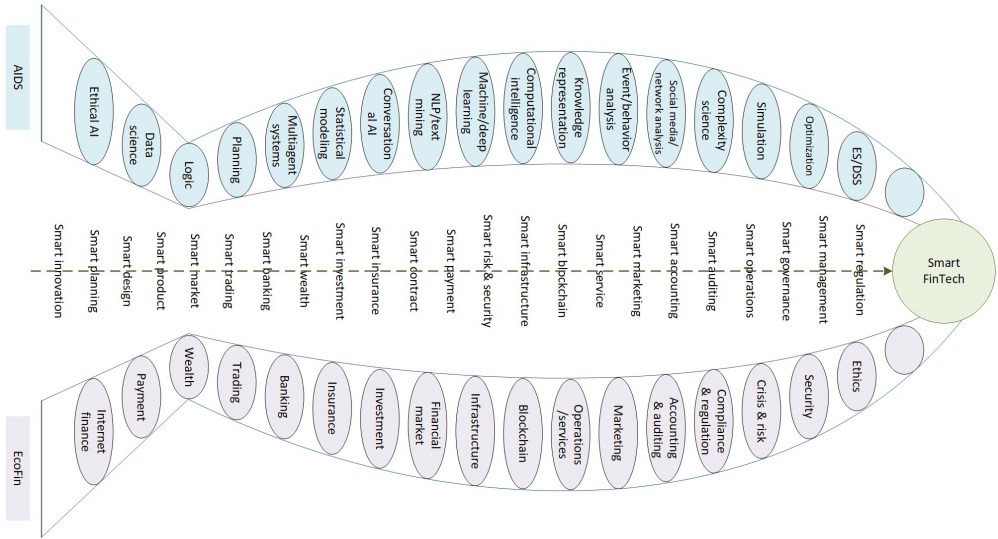

Fig. 1. Smart FinTech: The Synthesis between AIDS Techniques and EcoFin Businesses. The upper fin of the symbiont FinTech fish refers to the broad AIDS areas and techniques, the lower fin consists of major economic-financial areas and problems, and the ridge refers to the various synthetic areas of the smart FinTech family.

## 2.1 Economic-Financial Businesses

The EcoFin and FinTech business areas that build on and benefit from AIDS techniques involve not only almost all aspects of an EcoFin system and its environment but also broadly all EcoFin businesses [14, 19, 64, 90, 119, 121]. Different categorization methods can be used to define and group the EcoFin aspects and businesses, such as the business areas of FinTech introduced in Fig. 1 and the types of financial products and services [21, 90]. Below, we summarize and categorize the comprehensive EcoFin businesses in terms of typical financial products and services, the major procedural aspects of the EcoFin ecosystem or a specific market, and systematic alternatives. Despite the different highlights in each categorization and their potential categorization biases (interested readers may refer to specific references for other professional or focused structures), they all involve or produce a large quantity of objects, entities, their interactions and activities, and large amounts of lifetime data. Each of them is not only closely relevant to AIDS but also already attracts much research and innovation interest in the broad AIDS and EcoFin communities.

First, in finance, the main financial assets, products, instruments and their related operations and services that can benefit from AIDS include the following. (1) Stock and services: referring to stocks, bonds and other securities, and their relevant supporting operations and services. (2) Derivative and services: referring to underlying assets-derived instruments, e.g., futures, options, exchange-traded derivatives, Over-the-counter (OTC) derivatives (e.g., forwards and swaps), and their supporting operations and services. (3) Commodities and services: referring to hard and soft tangible commodities such as metals, diamonds (e.g., gold), energy (e.g., oil) and agricultural goods and electronic commodities (e.g., futures contracts, forward contracts, swaps, exchange-traded commodities (ETC), OTC commodities derivatives, and contract for difference (CFD) etc. derivatives), and their supporting operations and services. (4) Index and services: referring to hypothetical portfolios to measure market value, performance and characteristics, such as Dow Jones Industrial Average, S&P 500, Nasdaq Composite, FTSE 100, and bond market index. (5) Currency, cryptocurrency and services: referring to foreign exchange markets, cryptographic currencies (e.g., bitcoins), and their supporting operations and services. (6) Banking and services: referring to retail banking (incl. online, credit, saving and cheque), business banking, commercial banking, as well as loans and insurance, etc. derived banking services, and their supporting operations and services. (7) Insurance and services: referring to financial protection for auto, gap, health, income, casualty, life, property, liability, and credit, etc., and their supporting operations and services. (8) Wealth and services: referring to financial planning, investment management, and aggregating accounting and tax, etc. financial services to manage money, estate and other tangible and intangible assets, and their supporting services. (9) Surveillance and compliance: referring to accrediting institutions, laws, standards, policies and tools for enforcing, validating and managing governance, operation and regulation and their objectives, e.g., integrity, transparency, fairness, and efficiency.

Second, we categorize 15 high-level procedural aspects and areas universally embedded in both the broad EcoFin systems and the lines of businesses, processes and mechanics of a specific market. (1) EcoFin innovations: such as designing novel market mechanisms, financial products, payment methods, and IoT services. (2) EcoFin markets and mechanisms: such as physical or virtual organizations, the products and services traded in capital markets, e-commerce markets or energy markets; and their mechanisms and trading rules, e.g., limit order market mechanisms, business models for cryptocurrencies, and the underlying-derivative market connectivity. (3) EcoFin participants: such as individual and retail investors, institutional investors, service providers, and regulators. (4) EcoFin services: the services offered in a financial organization or market to fulfil its functions and value, e.g., retail banking, car insurance, peer-to-peer lending, financing, and online crowdfunding. (5) EcoFin valuation and pricing: such as the value estimation and pricing modeling of marketable securities, credit, properties, intangible assets, liabilities, and other goods and services. (6) EcoFin trading: including the systems, processes and rules to enable financing, investment, order execution or cancellation, etc. (7) EcoFin payment: including payment systems for a market, a company's finance, and the systems and services to enable online, mobile or contactless payment. (8) EcoFin systems and infrastructure: such as a stock exchange's trading systems, the fundamental trading and operational supporting systems, and blockchain-based distributed accounting systems. (9) EcoFin events and behaviors: such as investors' trading behaviors, market movements, company announcements, company mergers, and financial crisis. (10) EcoFin marketing and relationship management: including the activities and communications with stakeholders (e.g., for market campaigns), and maintaining and improving customer care, stakeholder relations, and business partnership. (11) EcoFin operations and resource management: including enabling the processes, facilities and services of a market, supporting financial innovations, design, production and services, and managing human resources, materials and intangible assets (such as data, information, patent and trademarks). (12) EcoFin governance, risk and compliance: including assuring organization's/market's objectives, operational orders and business rules for integrity, efficiency and fairness, managing risk, uncertainty and crisis on itself and stakeholders such as investment risk and enterprise/systemic risk. (13) EcoFin regulation: including enforcing regulatory laws, policies and regulations by independent regulatory and auditing authorities. (14) EcoFin security: including EcoFin system security, information security, and cybersecurity. (15) EcoFin ethics: including addressing social, political and ethical issues and privacy.

Lastly, we offer some systematic, cross-aspect, whole-of-view, hierarchical and spectral perspectives and categorizations to spotlight cross-system, cross-market, cross-participant, whole-of-enterprise, whole-of-business and whole-of-process spectrum, opportunities and areas for explorations. (1) Whole-of-business: such as connecting all products, services and businesses offered by an enterprise or a market for enterprise-wide and whole-of-business explorations, e.g., across the banking, trading, lending, insurance, wealth, marketing and payment businesses offered in a comprehensive bank. (2) Whole-of-operations: such as connecting all operations and functions over the entire lifespan of a market or organization, e.g., across the design, valuation, production, sales and marketing businesses of a financial service. (3) Cross-participant and entity hierarchy: such as connecting customers in different markets, products and services and exploring individual, group, sectoral participants, products and services. (4) Lifetime: such as fusing historical, present and future services and connecting static, dynamic, sequential and real-time EcoFin events, behaviors and activities. (5) Landscape: such as fusing individual, institutional, international and virtual (online) aspects for opportunity, risk, compliance, crisis and security explorations. (6) Business operational and support intelligence: including automating and personalizing design, pricing, packaging, financial, marketing and decision-support policies, processes, systems, services and management. (7) Cross-social, economic, ethical and political objectives and aspects: such as exploring their interactions and impact on a specific FinTech or market. Though these have not been the focus of the current FinTech research, they have received increasing attention in FinTech innovation [14, 33, 136]). These systematic FinTech perspectives and viewpoints need to be further explored for both holistic and specific developments and applications of AIDS techniques for smart FinTech.

## 2.2 Economic-Financial Business Challenges

The aforementioned EcoFin and FinTech businesses present many research opportunities and challenges to AIDS. They are related to not only fundamental business systems, processes, operations, regulations and management but also their planning, decision-making, monitoring and optimization. Such opportunities and challenges include but are not limited to mechanism design and optimization, forecasting and prediction, portfolio planning and optimization, sales and marketing analysis, business profiling, sentiment and intention modeling, anomaly detection, compliance enhancement, risk management, objective optimization, and operations optimization. We briefly discuss these below.

(1) Mechanism design and optimization: including designing, simulating, validating and optimizing market mechanisms for a market, product or service, e.g., the business models, pricing and stakeholder relationship models of a novel cryptocurrency. (2) Forecasting and prediction: including the regression, classification, estimation and prediction of trend (up or down), movement (direction and scale, etc.), value (e.g., price or volatility), temporal or sequential change at the next moment or interval, etc. (3) Portfolio planning and optimization: including designing, planning, optimizing and recommending investment portfolios and strategies in a market, for a product or entity, or across multiple markets (or financial variables). (4) Sales and marketing analysis: including characterizing, analyzing, evaluating, optimizing and recommending target products, markets and customers, sales strategies, supply and demand relationship, marketing campaigns, and customer relationships, etc. (5) Business profiling: including describing, segmenting, characterizing and classifying markets, products, customers, and services. (6) Sentiment and intention modeling: including characterizing, representing, modeling, analyzing and evaluating the polarity, diversity, propensity and their dynamics of customer sentiment and intention that may be associated with a market, product, institution, or participant type. (7) Anomaly detection: such as characterizing, quantifying, detecting, classifying and predicting abnormal, exceptional and changing behaviors, products, patterns, performance, relationships and structures, etc. associated with a market, product, institution or participant. (8) Compliance enhancement: such as characterizing, identifying, analyzing, categorizing and predicting compliance issues, scenarios, behaviors and their causing factors and evolution in a market, product, institution, participant or service; quantifying their consequences and impacts; and monitoring and improving the dynamic relations between incompliant behaviors, effects and compliance mitigation. (9) Risk management: including quantifying, analyzing, detecting, profiling and categorizing risk factors, risk areas, risk severity and consequences of risk in a market, product, institution, participant or service; recommending the corresponding risk mitigation strategies; and monitoring and improving the dynamic relations between risky scenarios, sequential risk, their effects and sequential risk mitigation. (10) Objective optimization: including identifying business objectives to be evaluated, balanced or optimized in a market, product or service; recommending strategies to balance or optimize the objectives; and evaluating the optimization effects, etc. (11) Operations optimization: such as detecting issues in business operations, governance and management of a market, product, service or institution; recommending the corresponding treatment strategies and plans and scheduling; and evaluating and optimizing the business and operational performance, etc.

## 3   ECONOMIC-FINANCIAL DATA AND CHALLENGES

The AIDS research in finance heavily relies on their data availability with a typical focus on understanding their data characteristics, challenges and potential for improving decisions, operations and management. This section discusses the various sources of EcoFin data and their challenges in relation to the corresponding AIDS research.

### 3.1   Economic-Financial Data

The EcoFin data and repositories include both internal and external sources. Many types of resources may be involved in AIDS-driven finance and economics research. Here, we categorize them into the following data types and briefly explain them: micro-EcoFin transactions, macro-EcoFin data, client data, operational data, EcoFin events and behaviors, EcoFin news and announcements, EcoFin reports, EcoFin social media and messaging, EcoFin cognitive data, accounting, taxation and auditing data, EcoFin feedback and question/answering data, simulation data, and third-party data.

(1) Micro-EcoFin transactions: including the micro-level transactions of an underlying EcoFin business, e.g., the trading transactions of an investor in a bond market, which often involve financial products, service time, actions, and other attributes (e.g., price and volume). (2) Macro-EcoFin data: including macro-level EcoFin transactions and data of a macro-product or indicator, e.g., GDP values, CPI, employment rates and petrol prices of a country in a year. (3) Client data: including the description of clients (consumers) of a product or service, e.g., investors' demographics in a foreign exchange market. (4) Operational data: including the description and recording of operations and management of an EcoFin business, e.g., business specifications, system settings, security, management and monitoring logs for operating and managing a product or service. (5) EcoFin events and behaviors: including actions and activities and their sequential developments undertaken by or resulting from a product or service, which could be micro, meso or macro, internal or external, or routine or exceptional, e.g., participant investment activities, natural disasters, political events, and trading manipulations. (6) EcoFin news and announcements: such as a press release from the media, communications about a new release or news about an accident involving a product, service or its institution. (7) EcoFin reports: such as formal EcoFin statements about the positioning, market activities, finance, and accidents involving a product, service, institution or participants, e.g., review reports, auditing reports, balance sheets, cash flow statements, income statements, and statements of equity and liquidity. (8) EcoFin social media and messaging data: including information communicated through social media or instant messaging channels (e.g., mobile apps) about a product, service, institution or participants, e.g., about the abnormal movement of a security price or the announcement of a stakeholder change of a listed company. (9) EcoFin cognitive data: about neural/brain activities and imaging, psychological and sentimental states and responses related to a product, service or participants, e.g., extracted from social media or customer service interactions. (10) Accounting, taxation and auditing data: that is related to a market, product, service or participants. (11) EcoFin feedback and question/answering data: collected from call centers, over service counters, physical or online interviews and questionnaires about a company, product or service. (12) Simulation data: collected from simulations about the functionalities, behaviors and performance of a market, product or service, e.g., the data collected from an artificial cryptocurrency simulation system or the testbed of a new product listing. (13) Third-party data: collected by third parties about an underlying product, service, institution or participants, e.g., the Bloomberg event-driven feeds, or data about relevant third-party products, services, institutions, or participants.

## 3.2 Economic-Financial Data Challenges

The above EcoFin businesses and data are coupled with each other in reality. This poses various opportunities and challenges for data-driven AIDS research [30, 31] in finance and economics. Here, we categorize them into the following perspectives that synergize EcoFin businesses and their data. (1) Innovation challenges: e.g., AIDS techniques for inventing novel, efficient, intelligent and sustainable mechanisms, products, services and platforms. (2) Business complexities: such as AIDS techniques for representing, learning and managing the intricate working mechanisms, structures, interactions, relations, hierarchy, scale, dynamics, anomaly, uncertainty, emergence and exceptions associated with a market, a product or participants. (3) Organizational and operational complexities: such as AIDS techniques to characterize and improve the diversity and personalized services of individual customers and sectoral demands, the departmental and institutional coherence and consensus in operations and services, and the inconsistent and volatile efficiency and performance in organization and operations. (4) Human and social complexities: such as AIDS techniques for modeling and managing the diversity and inconsistency of participants' cognitive, emotional and technical capabilities and performance and for enabling effective communications, cooperation and collaboration within a department and between stakeholders. (5) Environmental complexities: such as AIDS techniques for modeling and managing the interactions with contextual and environmental factors and systems and their influence on a target business system and its problems. (6) Regional and global challenges: such as understanding and managing the relations between an economy entity and its financial systems with the related regional and global counterparts and stakeholders and their influence on the target problems. (7) Data complexities: such as extracting, representing, analyzing and managing data quality issues, misinformation and complicated data characteristics, e.g., uncertainty, extremely high dimensionality, sparsity, skewness, asymmetry, and heterogeneity and couplings (i.e. non-IIDness) [27, 41]. (8) Dynamic complexities: such as modeling, predicting and managing evolving but nonstationary behaviors, events and activities of individual and batch markets, products, services and participants. (9) Integrative complexities: e.g., systematically modeling and managing the various aspects of the above complexities that are often tightly and loosely coupled with each other in an underlying EcoFin system.

In conclusion, the EcoFin businesses, data and their challenges discussed in Sections 2 and 3 pose numerous opportunities to the AIDS communities and smart FinTech. Below, we focus on reviewing the related techniques for data-driven AIDS research in finance and economics. This review complements the one in [33] that mainly takes a business application perspective.

## 4 AN OVERVIEW OF AI RESEARCH IN FINANCE

The AIDS techniques to support the aforementioned EcoFin businesses and process their data are very comprehensive, diversified and evolving. Such techniques address various aspects of business needs and problems, as reviewed in [33]. Here, we categorize the main AIDS techniques for smart FinTech into the following groups and briefly summarize their relevant work. Fig. 2 shows the overall classification of AIDS in finance. (a) Mathematical and statistical modeling: including numerical methods, time-series and signal analysis, statistical learning, and specifically random methods. (b) Complex system methods: including complexity science, game theory, agent-based modeling (ABM), and network science. (c) Classic analysis and learning methods: including pattern mining methods, kernel learning methods, event and behavior analysis, model-based methods, document analysis and NLP, and social network analysis. (d) Computational intelligence methods: including neural computing methods, evolutionary computing, and fuzzy set methods. (e) Modern analytics and learning methods: including advanced representation learning, optimization methods, reinforcement learning systems, deep learning systems, and visualization and interpretation techniques. (f) Hybrid methods: including ensemble methods and multi-method integration. Table 1 summarizes the AIDS techniques and their applications in EcoFin.

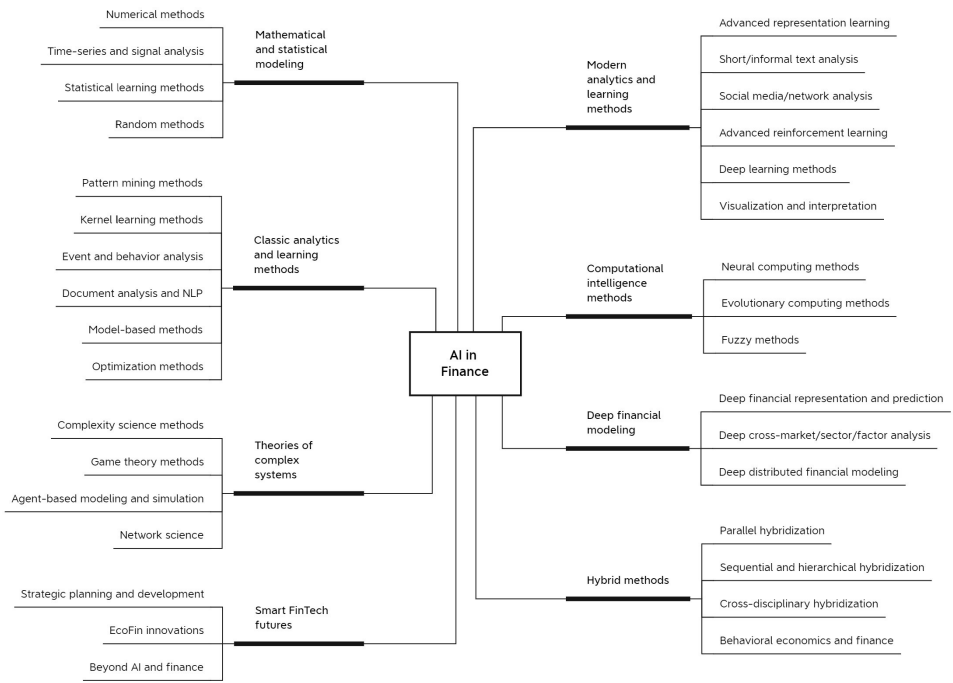

Fig. 2. The Technical Family of AI and Data Science in Finance.

## 4.1 Mathematical and Statistical Modeling

The families of mathematical and statistical modeling lay the foundation for characterizing, formulating and modeling EcoFin systems and their working mechanisms, problems and solutions [3, 153]. Accordingly, here we discuss four relevant techniques that are widely used to quantify and analyze EcoFin systems.

(1) *Numerical methods* are used to build quantitative representation and analysis of EcoFin systems and problems, e.g., for VaR, option valuation, option pricing, portfolio simulation, portfolio optimisation, hedging, and capital budgeting [77]. Typical methods include linear and nonlinear equations, least squares methods, interpolation, optimization, binomial and trinomial methods, finite difference methods, diffusion models, dependence modeling, financial simulation, Monte-Carlo simulation, random number generators, and econometric models such as term structure modelling and regression [77, 106, 142, 177, 197].

Table 1. AIDS Techniques and Their Representative Applications in Finance.

| AIDS areas | AIDS methods | Applicable EcoFin problems |
|---|---|---|
| Mathematical and statistical modeling | Numerical methods | Valuation, pricing, portfolio simulation and optimization, capital budgeting, hedging |
| | Time-series and signal analysis | Price prediction, market movement, IPO prediction, equity-derivative correlation analysis |
| | Statistical learning methods | Price estimation, value-at-risk (VaR) forecasting, financial variable dependency modeling, portfolio performance estimate |
| | Random methods | Abnormal behavior analysis, market event analysis, influence transition analysis, associated account analysis |
| Complex system methods | Complexity science methods | Market simulation, mechanism design, globalization analysis, crisis contagion, market information flow |
| | Game theory methods | Policy simulation, regional conflict, mechanism testing, and cryptocurrency mechanism testing |
| | Agent-based modeling | Testing economic hypotheses, simulating policies, supply/chain relations, portfolio optimization |
| | Network science | Modeling entity movement, community formation, interactions and linkage, influence and contagion propagation |
| Classic analytics and learning methods | Pattern mining methods | Trading behavior analysis, abnormal trading, outlier detection, investor relation analysis |
| | Kernel learning methods | Price and market movement prediction, cross-market analysis, financial crisis analysis, crowdfunding estimate |
| | Event and behavior analysis | Financial event analysis, price co-movement, abnormal behavior analysis, market event detection |
| | Document analysis, text mining, and NLP | Financial event analysis, sentiment analysis, company valuation, financial reporting and review, auditing, fake news and misinformation analysis |
| | Model-based methods | Hypothesis testing*, market index modeling, event analysis, fraud detection, movement forecasting |
| | Social media and network analysis | Social influence analysis, sentiment analysis, opinion modeling, customer feedback analysis, market and price movement, associate account detection, detecting manipulation, and insider trading |
| Computational intelligence methods | Neural computing methods | Macroeconomic and microeconomic factor correlation, valuation, portfolio optimization |
| | Evolutionary computing methods | New product simulation, financial objective optimization, market performance optimization |
| | Fuzzy set methods | Modeling market momentum, financial solvency, risk and capital costs |
| Modern AIDS methods | Representation learning | Representation of stocks, assets, markets, portfolios, events, behaviors, and financial reports |
| | Short and informal text analysis | Text-based trend forecasting of price, market, sentiment and reputation, question/answering |
| | Optimization methods | Optimizing policies, portfolios, trading strategies, VaR, and market performance |
| | Reinforcement learning methods | Simulating and optimizing supply/demand of new assets and services, discovering trading signals, portfolios and investment actions, optimizing portfolios and trading strategies |
| | Deep learning methods | Market modeling, behavior modeling, trading modeling, risk analysis, price and movement prediction |
| Hybrid AIDS methods | Parallel ensemble | Price and market movement forecasting, risk analysis, financial event detection, customer profiling. |
| | Sequential and hierarchical hybridization | Financial review-based fraud detection, macroeconomic influence on market movement, social media impact on price movement, epidemic evolution and impact on market volatility, market trend and confidence |
| | Cross-disciplinary hybridization | Psychological factors and irrational market behaviors, behavioral economics and finance, sentiment and intention modeling, misinformation and mispricing on market inefficiency |

(2) *Time-series and signal analysis* describe, characterize, analyze and forecast temporal movement of financial variables, their state transition, regression, trends and changes by treating EcoFin indicators as time-series and signals [11, 55, 160, 167]. Examples are analyzing the signals and their relations of a security price and its derivative price and their market index by multivariate regression and modeling high-frequency trading signals jointly by time-domain sequence analysis and frequency-domain wavelet. Typical methods include state space modeling, linear time-series analysis, nonlinear time-series analysis, time-frequency and time-scale analysis, spectral analysis, Kalman filter, fractional time-series analysis, long-memory time-series analysis, seasonal time-series analysis, transfer function models, multiscale analysis, multivariate analysis, stationary analysis, nonstationary analysis, and outlier analysis [5, 11, 55, 75, 128, 161, 167, 199].

(3) *Statistical learning methods* measure, estimate and learn the uncertainty, randomness, risk, pricing, rating, performance or dependence of EcoFin systems, products and problems in terms of probabilistic theories [3, 153, 175]. Examples are estimating the pricing of options, forecasting the VaR and performance of a portfolio, modeling the sequential trading behaviors of a group of investors or a firm by coupled hidden Markov models, modeling the high-dimensional dependencies between multiple time series by a copula method, and measuring the probabilistic relations between financial indicators as a Bayesian network. Typical methods include random walk models, efficient portfolio theories, factor models, Black-Sholes models, Monte-Carlo methods, Delta-Gamma approximation, Le´vy processes, stochastic volatility models, copula methods, filtering methods e.g. particle filters, and nonparametric methods such as Bayesian networks and Markov networks [23, 36, 41, 48, 56, 152, 153, 169, 175, 198, 206].

(4) *Random methods* characterize, model, simulate and analyze an EcoFin problem in terms of the theory of randomness and uncertainty [82]. Examples are modeling the evolution of abnormal and exceptional market behaviors, global events, and black swan events. Such methods include random sampling, random walk models, random forest, stochastic theory, fuzzy set theory, and quantum mechanics [102, 108, 122, 140].

## 4.2 Theories of Complex Systems

The theory of complex systems has been widely used in classic AI and applied in EcoFin for understanding, simulating and analyzing working mechanisms, system characteristics and complexities, and the emergence and consequences of EcoFin systems and problems [29, 188]. We introduce four such methods below.

(5) *Complexity science methods* model an EcoFin system (e.g., a bitcoin market) as a complex system and understand its intrinsic and intricate mechanisms and complexities, global economy and its evolution, inter-regional and inter-country relation analysis, migration, crisis contagion, conflict modeling, international trading and information flow. Combined with sociology and systems theory, typical methods include theories of complexities, systems theory, emergence, self-organization, complex adaptive systems, ABM, chaos theory, and random fractal theory [61, 176].

(6) *Game theory methods* build mathematical models to design, characterize, simulate and analyze the interactions, conflict, cooperation, communication, coalition, uncertainty, behavioral relations, social norm, reputation, trust, Nash equilibrium, and consensus-building mechanisms and processes in complex EcoFin systems and to design strategies for market mechanisms, pricing, bargain, contracting, corporate finance, accounting, and information asymmetry, sharing and transparency. They can be used to model the conflict between political systems, estimate rational and irrational threats in regional conflict, and model the market mechanisms of blockchain and cryptocurrencies by continuous game theories, to name a few. Typical methods include zero-sum games, continuous games, differential games, combinatorial games, evolutionary games, stochastic games, Bayesian games, strategic-form games, extensive-form games, and the communication, bargaining, cooperation, coalition in collective and cooperative games [45, 132, 173].

(7) *Agent-based modeling and simulation* models, simulates and optimizes the working mechanisms, dynamic formation processes, internal interactions between entities (i.e., agents), and understands and validates the working mechanisms and evolutionary processes of EcoFin systems. ABM builds multiagent systems to simulate an EcoFin problem as a complex system, and incorporates mechanisms including perception, interaction, policy selection, rules, reinforcement learning and optimization into simulating, optimizing and validating the working mechanisms in such systems. ABM has been widely explored in economics and finance [35, 53, 66, 81, 113, 189].

(8) *Network science* characterizes, models, analyzes and predicts the directed and undirected connections and interactions between participants, products and actor behaviors, entity movement, community formation and influence propagation and contagion in EcoFin systems by modeling such systems as complex social or information networks. Typical methods include network linkage analysis, graph networks, scale-free and power law, small worlds, influence diffusion and contagion theories and tools [73, 83, 97, 98, 171, 180]. Such methods can be used to model group investor interactions, relations and communications as a directed network e.g. by Poisson factorization-based Bayesian models, model the interactions between market participants in an economic market, and model problems in EcoFin systems e.g. corruption risk in contracting markets [187].

## 4.3 Classic Analytics and Learning Methods

Classic analytics and learning methods have played critical roles in making EcoFin intelligent by analyzing EcoFin data, and discovering and optimizing the patterns, clusters, classes, trends and outliers in EcoFin systems [31]. We categorize them into the following six methods: pattern mining methods, kernel learning methods, event and behavior analysis, document analysis and NLP, model-based methods, and optimization methods. Readers can also refer to [33] for more discussion on the business applications of classic research topics in EcoFin and many other references on classic analytics and learning methods.

(9) *Pattern mining methods* model and discover patterns and patternable behaviors, structures and relations in EcoFin systems. Examples are identifying arbitrage trading behaviors by mining frequent and high-utility cross-market investment strategies, discovering high-frequency trading strategies, detecting periodical price or market movements, and analyzing financial and social relations between investors and firms. Typical methods include discovering frequent patterns, sequential patterns, graph patterns, network patterns, tree patterns, knot patterns, interactive patterns, combined patterns, and abnormal trading behaviors [38, 63, 103].

(10) *Kernel learning methods* describe, represent and analyze the distributions, numerical relations and similarity between EcoFin indicators (e.g., markets, securities, or firms) by individual or multiple (linear or nonlinear) kernel functions and their couplings. Typical methods include linear and nonlinear kernels, vector (space) kernels, tree kernels, sequence kernels, support vector machines, Fisher kernels, spectrum kernels, multi-kernel learning and graphic kernels [103, 118, 166]. Such methods can characterize the inter-dependencies between micro-market variables such as security price and volume, between macro-economic indicators such as GDP values, exchange rates and gold prices, and between financial texts by multi-distributional kernel functions.

(11) *Event and behavior analysis* characterizes the occurrences, driving forces, evolution and consequences of events, activities and behaviors [216] undertaken by or on an EcoFin object and detects and predicts abnormal, unexpected and changing events and behaviors [36, 37, 68]. Examples are detecting exceptional co-movements between an underlying equity, derivative products, and the equity company's announcement releasing activities by coupling learning and coupled behavior analysis of their activity sequences. Many other applications exist, e.g., analyzing group/pool manipulation [36], cross-market behavior couplings [41], market herding behaviors [215], and information security events [170]. Typical methods including historical event analysis, sequence analysis, Markov chain process, nonoccurring behavior analysis, high-impact behavior analysis, high-utility behavior analysis, and group behavior analysis [169], and deep recurrent neural networks [99].

(12) *Document analysis and NLP methods* recognize, identify, extract, summarize, classify, search and compare rule-violated, suspicious, concerned and risky statements, claims, announcements, price and market movement, concepts and topics, emotions and sentiments, and entities and misinformation etc. within and between EcoFin documents, reports, and news [96, 130, 135, 165, 203]. Examples are identifying and comparing problematic descriptions, revenue and budget statistics, misleading or misclassified reporting and review comments across monthly financial review reports by Transformer and BERT-derived neural models [59]. Typical methods include language models, named entity analysis, case-based reasoning, sequence labeling, statistical language models, context-free parsing, logical and dependency semantics, distributional and distributed semantics, topic modeling such as latent Dirichlet allocation, word embedding, Bayesian networks, and neural models such as Transformer, BERT and their variants.

(13) *Model-based methods* characterize, represent and analyze EcoFin phenomena, events, problems and dynamics in terms of the given hypotheses and models. Examples are modeling trading behaviors, market and price movements, institutional trading behaviors, price and index movement trends, and the influence of traders' sentiments on the dynamics of financial mechanisms and markets. Typical models include numerical computation models such as kernel functions, statistical models such as expectation-maximization models (e.g., Gaussian mixture) and probabilistic graphical models (e.g., Bayesian networks and Markov networks), machine learning models such as for clustering, classification and semi-supervised learning, deep reinforcement learning (DRL) models, and deep neural networks [15, 23, 41, 44, 56, 82, 109, 152, 156, 186].

(14) *Optimization methods* model EcoFin systems and problems as optimization problems or apply optimization methods to characterize, analyze and recommend optimal EcoFin solutions. Examples include optimizing portfolio design and strategies and return on investment w.r.t. relevant micro-and macro-level financial indicators; optimizing the VaR of a cross-market portfolio with transaction cost; and optimizing algorithmic trading strategies [8, 47, 51, 77, 105, 131, 138, 145]. Typical methods including classic numerical methods such as linear programming, nonlinear programming, quadratic programming, stochastic programming and dynamic programming and advancements in the aforementioned genetic programming, multi-objective evolutionary computing, swarm intelligence, information theory, Bayesian optimization, machine learning methods, and DNNs-based methods.

### 4.4   Computational Intelligence Methods

Computational intelligence methods model the working mechanisms of EcoFin systems, investment analysis, economic forecasting, portfolio analysis, inflation prediction etc. by computing mechanisms inspired by natural, biological and evolutionary systems and fuzzy logic.

(15) *Neural computing methods* model the relations, structures, sequential movements, transitions and influence between EcoFin variables by neural networks, in particular, deep neural networks (DNN) nowadays [179]. Examples are modeling the dependencies between a stock price, market index, foreign currency and derivative market movement and macroeconomic variables such as petrol and gold prices; detecting abnormal market behaviors, anomalies or frauds in corporate finance, accounting, insurance and banking businesses; predicting corporate bankruptcy and failures; predicting pricing and hedging strategies and algorithmic trading strategies; and analyzing financial reports, announcements and events [16, 85, 89, 92, 104, 115, 139, 144, 154, 161, 179, 201, 212]. Typical methods include various classic artificial neural networks (ANN), recurrent neural networks (RNN), wavelet neural networks, genetic neural networks, fuzzy neural networks, and DNN variants including attention networks and Bayesian neural networks.

(16) *Evolutionary computing methods* characterize, simulate, analyze and optimize the working mechanisms, evolution, interactions, variances, performance and risk of EcoFin systems inspired by biological and evolutionary systems and models [146]. Examples are characterizing, simulating and optimizing the development and evolution of a newly released financial products in a market by a particle swarm organization (PSO) model; estimating and optimizing security price, market index, exchange rate and inflation rate; discovering trading rules for algorithmic trading; detecting operational issues; supporting credit scoring and profiling; and multiobjective optimization in EcoFin [86, 105, 123, 147, 150]. Typical methods include ant algorithms, self-organizing map, genetic computing/programming, artificial immune systems, gene expression programming, particle swarm optimization, swarm intelligence, and neural-genetic algorithms.

(17) *Fuzzy methods* characterize the relationships, distributions and structures of EcoFin systems in terms of fuzzy set and logic theories and fuzzy systems [122]. Examples are modeling market momentum, price, capital cost, risk, financial solvency, financial structures, the relations between costs and profit and between financial structures and capital costs, and multiobjective evolution [26, 80, 122, 193]. Typical methods include basic fuzzy approaches built on fuzzy set theories, fuzzy logic and fuzzy systems and hybrid methods integrating other models such as fuzzy neural networks, genetic fuzzy logic models, fuzzy convolutional networks, and fuzzy rough neural networks.

### 4.5   Modern Analytics and Learning Methods

Several categories of modern AIDS methods, including the new-generation developments of classic approaches for representation learning, short and informal text analysis, social media/network analysis, reinforcement learning, and deep learning, have been the focus in the recent decade or so for EcoFin problem-solving. Readers are also encouraged to refer to [33] and many specific surveys and technical papers for more discussion on the state-of-the-art AIDS research topics and progress in EcoFin.

(18) *Representation learning methods* describe, characterize and model the intrinsic processes, interactions, relations, structures, distributions and characteristics of EcoFin systems, products or problems. Examples include establishing a probabilistic, mathematical, graph-based, network-based, tree-like or neural representation of the assets, participants, and role interactions in a derivative market; representing securities by involving tick-by-tick data and external news; and characterizing the investment sentiment in terms of the social media communications about a security. Typical methods include visual models (e.g., candlestick chart), probabilistic models (e.g., latent Dirichlet allocation for sentiment and topic representation), graph networks (e.g., Bayesian networks for representing price movement), interactive models (e.g., agent-based reinforcement learning), network embedding (e.g., neural networks-based embedding of investment behaviors), tree models (e.g., regular vine copula for dependence structures between financial variables), and neural embedding (e.g., graph neural networks for stock representation and attention network for contextual representation) methods [18, 49, 54, 57, 96, 107, 110, 191, 196, 218].

(19) *Short and informal text analyses* collect, extract, recognize, analyze and classify short texts in social media, SMS and instant messaging systems about EcoFin institutions, products, services, trends, news or participants. Such information is alternative to core data available in markets and EcoFin institutions but consists of important messages about the underlying objects. Financial short text can be used for various purposes, e.g., predicting the trend and movements of price, market index and exchange rate; extracting and representing keywords, phrases and expressions in a social and virtual community about manipulating a commodity price; predicting the market sentiment, reputation and confidence on a product or service; and characterizing the effectiveness and efficiency of and discovering issues and solutions in customer communications through counter services, call centers and online chatbots, etc. [52, 96, 116, 130, 135, 165]. Typical methods include explicit short text understanding such as conceptualization for segmentation and labeling, term/tag/phrase similarity learning, dependency parsing, syntax structure analysis, short text classification, query and recommendation; and implicit short-text analysis, e.g., word/phrase/sentence embedding, contextual embedding, short-text conversation (e.g., chatbot), question/answering by neural models such as long short-term memory, attentive RNN, neural models with multi-head attention, encoder-decoder, and variational autoencoder, Transformer and its variants, etc..

(20) *Social media and network analysis* characterizes, detects, classifies, clusters and predicts linkages, interactions, feedback, sentiment, networking behaviors and abnormal interactions (e.g., pool manipulation or insider trading) between entities of EcoFin systems. Such methods model EcoFin systems as social networks by involving networking data, messaging data, feedback and sentiment in the communications, etc. from relevant social media, mobile apps, and instant messaging platforms. They can be used to characterize, quantify and trace the co-integration between social communications and market trading behaviors, sentiment and opinion and their influence on a security price or market index movement by methods such as short text analysis, network science methods, and social media analysis. Typical methods include social media analysis, topic modeling, sentiment analysis, emotional analysis, influence analysis, modeling linkage and interactions, multimodal analysis, deep reinforcement learning, recurrent neural networks, and neural language models such as BERT and variants [6, 9, 54, 73, 96, 100, 134, 174, 213].

(21) *Reinforcement learning* models an EcoFin system as a reinforcement learning process by defining and optimizing actions, policies and rewards (value) [70, 72, 159, 202]. Examples include modeling the portfolio management and adaptive configuration of derivative products to optimize the trading objectives, portfolios and strategies for algorithmic trading; representing and forecasting financial signals (e.g., prices) and trends of securities, foreign currencies, and commodities; modeling the supply-demand relations and stakeholder relationships in property markets, energy markets and capital markets; optimizing order routing in foreign exchange and features markets; and optimizing trading agent actions, option valuation, trading and asset management [2, 22, 44, 57, 70, 72, 76, 110, 112, 117, 125, 158, 159, 202]. Typical methods include classic reinforcement learning methods such as on-policy learning e.g., Bellman Equation, on-policy actor-critic reinforcement learning and Markov dynamic process reinforcement learning and off-policy learning e.g., Q-learning and deterministic policy gradient; and DRL [159] such as deep Q-network (DQN), double DQN, $\epsilon$-greedy policy Q-learning, $\epsilon$-greedy policy double Q-learning and weighted Q-learning, deep deterministic policy gradient, hierarchical reinforcement learning, recurrent reinforcement learning by integrating RNN with Q-learning, and adversarial learning in reinforcement learning.

(22) *Deep learning methods* (here mainly referring to deep neural networks) represent financial variables and their hidden relations and structures in EcoFin systems by deep abstractions and representations [79, 82]. Deep learning can be applied for the representation of stocks, assets and portfolios described by their financial indicators and variables; modeling EcoFin signals, trends, movements, events, behaviors and their sequential developments; developing trading strategies over sequential trading; predicting the risk, profit, return and VaR of portfolios; characterizing and analyzing the couplings, interactions and connections between macroeconomic and microeconomic variables, across markets, across participants and multi-product portfolios; detecting abnormal signals, movements, behaviors, events and interactions in EcoFin systems and networks; discovering incompliant behaviors and events in financial reports, news and announcements; understanding customer sentiment and feedback driving their EcoFin activities for tailored investment education and marketing; supporting smart crowdfunding and crowdsourcing; modeling the influence of external events, news and parties on the target markets, variables or objectives; analyzing and predicting the likelihood of financial crisis, vulnerability, and cross-market influence; and detecting financial misinformation and criminal activities (such as money-laundering and market manipulation) [40, 54, 79, 82, 141, 151, 159, 192]. Typical deep models including convolutional neural networks (CNN), RNN, attention networks, memory networks, generative adversarial networks (GAN), encoder-decoder, autoencoder, graph neural networks (GNN), deep Bayesian networks and neural language models (e.g., Transformer) can be used for the above EcoFin purposes. More discussion is available in Section 5.5 on deep financial modeling.

## 4.6 Hybrid AIDS Methods

Many EcoFin systems and problems have to be handled by hybridizing, ensembling or synthesizing multiple complementary AIDS techniques. Such hybrid and ensemble approaches are typically used to address (a) single yet complex EcoFin problems that cannot be well handled by single models; and (b) compound problems that require multiple methods to address various challenges. Accordingly, there are two main directions of synergy. Accordingly, the aforementioned AIDS techniques may be selectively integrated in parallel, sequential or hierarchical ways to enhance their capacity and ability. Typical hybrid methods can be categorized into three types: parallel hybridization, sequential hybridization, hierarchical hybridization, and cross-disciplinary hybridization.

(23) *Parallel hybridization*, similar to ensembling, integrates multiple same or different models to address the underlying problems. Examples include integrating expert systems and evolutionary computing for advising investment [150]; combining evolutionary computing and neural computing to generate evolutionary neural models to simulate and analyze a capital market [20]; integrating complexity science and game theory to model and simulate blockchain and cryptocurrencies [14, 87]; training a Bayesian network integrated with Bayesian dependence learning, neural networks or other machine learning mechanisms [178]; and learning a copula variational autoencoder or GNN to model dependencies, distributions, and connectivity in EcoFin [182].

(24) *Sequential and hierarchical hybridization* integrates multiple methods sequentially or hierarchically to address underlying EcoFin problems. Examples are to model stock price relations by time series analysis and then predict the abnormal movement by a classifier such as DNNs [124, 160, 162, 181, 217]; model the micro/macro-economic variable dependency using a copula method or GNNs and then forecast the VaR by a time series model [191, 206]; hierarchically represent market variables as input and then model the trading behaviors in a market by a RNN before the prediction of market movement. Similar parallel, sequential and hierarchical hybridizations of AIDS techniques can be widely seen in the relevant references.

(25) *Cross-disciplinary hybridization* integrates economic theories, financial theories, social science with AIDS methods, which have been booming in recent years. Examples include migrating multivariate time series to deep sequential models; combining copula methods with deep neural models; integrating autoregression with DNNs; learning high-utility behavior sequences; training actionable trading agents and strategies for learning trading history and satisfying portfolio optimization objectives (e.g., VaR); and fusing sentiment, trend and policy optimization to develop investment strategies [3, 16, 78, 88, 93, 99, 134, 151]. This direction presents a more genuine yet challenging approach for addressing significant EcoFin complexities.

(26) Lastly, *behavioral economics and finance* is a typical area that integrates multiple disciplines of thinking, theories and methods (e.g., psychology, neuroscience, applied behavior analysis, and computational intelligence) to understand, interpret and model the less rational to irrational investment, mispricing behaviors, market inefficiency and market hiddenness that drive market decisions and evolution. It migrates from the hypothetical rationality, bounded rationality and efficiency and transparency in classic economics and finance (corresponding to the 'System 1' mode of thinking and decision-making in psychology [95]) to the less rational and irrational human choices and market inefficiency (i.e., System 2) in real-life human decision-making. Typical behavioral economics and finance methods include prospect theory, nudge theory, natural experiments, and experimental economics [64, 71, 97, 99]. More advanced directions are to incorporate AIDS-driven modeling, analytics, simulation, learning and operation into behavioral economics and finance to understand, analyze and interpret human cognitive intent, emotion, thinking and decision-making in markets.

## 5 DATA-DRIVEN AI IN FINANCE

EcoFin is one of the areas with the richest data and where data-driven AI plays a critical role. To take advantage of data-driven AI for smart FinTech, it is necessar to tackle the characteristics and complexities of EcoFin data and extract the potential, intelligence and wisdom from the data for smart decision-support and making [7, 16, 30, 31, 44, 54, 63, 70, 188]. Here, we review the related research and group data-driven AI in EcoFin into five major families for review: EcoFin time series analysis, long and short EcoFin text analysis, EcoFin behavior and event analysis, multisource EcoFin data analysis, and deep financial modeling.

### 5.1 Economic-Financial Time Series Analysis

Time series are a typical data type and format in EcoFin businesses [23, 41, 55, 160, 167, 198, 201, 204, 206]. EcoFin time series consists of *macro-level time series* such as macroeconomic and social variables and indicators and *micro-level time series* such as asset prices, volumes, indices and values of assets such as securities, derivatives, index products and commodities, and foreign exchange rates. EcoFin time series present various time series characteristics, such as temporal, sequential, biased, tailed, stylish, dynamic, stochastic and uncertain characteristics. When *multiple time series* (or *multivariate time series*, MTS) are involved, they are often strongly and weakly coupled with each other across temporal intervals and at individual time points in terms of different coupling relationships and interactions such as uncertainty, dependency, correlation and causality [10, 28], though each time series of MTS may hold its own data characteristics such as distributions, structures and relations. Time series may also involve other value and numerical data, such as cash flow, property value, and monetary factors.

EcoFin time series has been widely analyzed for many purposes on various time series data. Examples include:

- Learning the representations of financial time series such as stock price and market index by deep autoregressive models, long-short term memory networks (LSTM) or CNNs. When labels are available (e.g., historical trends and performance in historical data), one can then train the representations by an encoder or embedding per mutual information or loss functions. The learned time-series representations can be used to further learn a regression of each time series for the prediction of its movements [163, 181, 217].
- Conducting tick-by-tick and high frequency price forecasting by modeling the historical time series of the prices as input, the converted sequence by RNN, and then predicting future prices by regression or RNNs [115, 151].
- Modeling stock price or market index movements by sequential processes such as RNNs, Hawkes processes, or dynamic Markov processes [36, 50].
- Predicting a stock market's index based on the efficient market hypothesis, time series analysis, ANN or DNNs of a market's index transactions [41, 201].
- Modeling the inter-financial variable couplings and influence by considering the couplings between variables, the dynamics of each time series, the dynamic couplings between MTS, the stylish patterns and variance in each variable and in the MTS couplings [39–41, 198]. For example, one can first model the coupled stock representations, then learn the trends of each stock by considering its couplings with others, and lastly use a reinforcement learning model to enhance the portfolio objective (e.g., per the value at risk or Sharpe ratio) [205, 206].
- Modeling financial MTS by quantum-based entropy time series [140], calculating the similarity between a new time series and the existing ones by dynamic time warping methods and modeling the co-evolving momentum and trends between time series.
- Modeling the dependencies between MTS by coupled hidden Markov models, probabilistic models or DNNs [36, 41, 137, 169], e.g., by RNNs where each model (or channel) captures the sequence of a time series and multiple models capture the interactions between the series.
- Conducting time series forecasting by transfer learning across multiple sources or modalities of data [93, 109, 133, 157], e.g., borrowing knowledge learned from an equity to model the price dynamics of its derivatives or cross-sectional stock investment [133, 218].
- Modeling the couplings and decouplings between time series [40, 198] and the changes and inconsistencies between time series for representing stocks and cryptocurrencies [5, 128].
- Modeling foreign exchange rates by time series and sequence analysis, ANN or DNNs and analyzing the relationships between multi-foreign currency rates as MTS [89, 115].

Classic techniques for EcoFin time series analysis include *time series analysis* methods such as regression, autocorrelation analysis, linear programming and trend forecasting methods [23, 55, 160]; and *signal processing techniques* such as spectral analysis, wavelet analysis, Hawkes processes and Markov processes [126, 169]. Recent years have seen more advanced time series analysis techniques, e.g., multiple sequence analysis, nonoccurring (negative) sequence analysis, and neural sequence modeling such as RNN [34, 191].

## 5.2   Long and Short Economic-Financial Text Analysis

In EcoFin, the sources of *long economic-financial text* include EcoFin news, announcements, financial reports, annual reports, project proposals, and statements [67, 92, 96, 127, 135, 165, 203]. The sources of *short economic-financial text* include communication and conversation data such as call centre data, recording, and chatbot manuscripts; social media data and blogs; short financial reports and bank statements; and SMS and instant messaging. Long and short EcoFin text may be both online and offline. For simplicity, we call any long and short EcoFin texts 'economic-financial text'.

EcoFin issues and objectives which benefit from analyzing EcoFin text include but are not limited to the extraction, categorization, classification and prediction of fraud, event, knowledge, sentiment, topics, opinion and stories about a financial company, instrument, product, or service [52, 96, 130, 135]. Such information can be used for different financial engineering, economic computing, quantitative analysis, data-driven analytics, and decision-support purposes and tasks. Some examples are as follows.

- Extracting well to weakly declared or undeclared but indicative information, opinion, sentiment and expert opinion about a company, financial instrument, product, service, or a sector [100, 134].
- Extracting knowledge and profile about an asset, a company, and a sector, etc. from mixed-format EcoFin text (e.g., text description with tables and statistics) [62].
- Analyzing, modeling and predicting the normal or abnormal price or index trend and movement of an asset at a time or during a time period based on the information and influence of financial text [207].
- Analyzing, predicting and interpreting the sensitivity and influence of EcoFin text on an asset price, its trend, movement or its market popularity and profitability, and predicting the asset price, movement and performance [135, 165].
- Analyzing, predicting and interpreting the sensitivity and influence of EcoFin text on a company's branding, reputation, competitiveness, and performance, etc. [130].
- Classifying and summarizing EcoFin text and its polarity on market movement [96].
- Building the embedding and representation of an asset, a listed company, a financial product or a service based on its related financial text and the context, e.g., the industry sector and market performance [92, 118].
- Extracting risk-sensitive factors, events, entities and sentiments in financial text about an asset, a company, a product or service; quantifying the risk rating and ranking of assets, companies, products or services in terms of the extracted risk-sensitive indicators [68, 130].
- Analyzing and identifying the concerned financial business areas, products, services, topics, events, and systems in marketing, sales and customer services-related customer communications and conversations, such as from call centre records, over-the-counter service recordings, online messaging such as chatbot manuscripts, Q/A websites, email communications, and instant messages in social media and mobile apps [9, 131].
- Extracting and quantifying the quality, performance, problems, questions, gaps and demand about financial products and services from contextual human conversations in Q/A, call centre records, chatbot and email communications, conducting discourse analysis and argumentation summarization, and making answer, service and product recommendations [74, 130, 174].
- Identifying and categorizing opinions, complaints, sentiment, emotion, concerns, feedback and expert rationale and opinions in social media data, instant messages, news, and discussion groups [6, 96].

AIDS techniques that can be used to analyze financial text include typical *text mining* and *document analysis* methods for document clustering, document classification, text summarization, text categorization, concept and entity extraction, and entity relation modeling such as concept models, statistical models such as Bayesian networks, decision trees, and case-based reasoning; *NLP* techniques such as part of speech tagging, syntactic parsing, and linguistic analysis; *topic modeling* and *sentiment analysis* methods such as latent Dirichlet allocation models; *word embedding* models such as Word2Vec, Glove and BERT; *RNNs* such as LSTM for sequence modeling; new *neural language models* such as Transformer and its variants; and *attention networks* for neural word embedding and language models (i.e., neural attention). The difference between applying the above techniques in long versus short financial texts lies in the hierarchy of content where long text may involve word to sentence, paragraph and document layers, while short text may mainly involve words and phrases; the length of input; and the context of input whether there are pre- and post-sentences and paragraphs; and learning tasks whether the learning target refers to individual words, next sentences, or a categorization of the input.

## 5.3 Modeling Economic-Financial Behaviors and Events

In EcoFin, *behaviors* refer to the investment activities conducted by investors, activities undertaken by market makers, actions taken by service providers and recipients, and actions taken by regulators, etc. More weakly and fuzzily annotated terms such as *trading behaviors* and *investment behaviors* refer to the trading actions placed on an asset in a market, e.g., placing a buy order on a stock; *service behaviors* refer to such activities conducted by a provider or recipient of an EcoFin service, e.g., overclaiming tax, repaying a debt, charging a credit card for online shopping, and suspending a compromised card; and *market behaviors* refer to the movement of a market, e.g., moving up or down. An EcoFin *event* refers to the occurrence of a thing or an occasion taking place in a financial company, by a regulator, a market management team or a policy-maker, or in other public spaces, e.g., a company announcement release, a change of interest rate or the occurrence of a bushfire. All individual, group and firm-oriented behaviors and events may be involved. Behaviors and events may be static or dynamic. *Dynamic behaviors* (or events) refer to their sequential behaviors which have taken place over time. AIDS techniques and behavior informatics [27, 36, 37] can be applied on EcoFin behavior and event data for different business problem-solving and analytical tasks [36, 37, 68, 99, 114, 211, 216].

- Analyzing investor's behaviors such as buys, sells, holds and withdrawals in investment transactions; constructing a sequence of investment behaviors; modeling and detecting high-performing, poor-performing and exceptional investment behavior patterns; and identifying tailed behaviors, erratic behaviors, and extreme behaviors [99, 215].
- Given multiple groups of behaviors, e.g., trading behaviors of different market makers or institutions, modeling the similarity and difference between their trading behavior sequences and patterns, detecting anomalies (e.g., pool manipulations, herding behaviors) across multiple behavior sequences, and improving marketing and customer services [74, 169].
- Given investment behaviors and market behaviors, modeling their influence on the movements of an asset, a market or across markets, and predicting the asset price trend, market movement, or cross-market movement [36, 42].
- Detecting, summarizing, classifying and interpreting EcoFin events in EcoFin text based on event triggers, event types, historical events, and interested or concerned topics and keywords; when there are no event-related labels available, conducting unsupervised financial event extraction from EcoFin text by learning the contextual representation of EcoFin documents and extracting event arguments [68, 118, 120, 170].
- Extracting financial events from EcoFin text by considering the writing styles, the layout features, and domain expert-given event types and categories, etc., in the text; and learning novel event types and categories based on the prior knowledge and weakly supervised information learned about events [127, 166].

Techniques for behavior modeling, behavior informatics, user modeling, event modeling, historical event analysis, event detection, positive (occurring) and negative (non-occurring) sequence analysis [34], sequence modeling, neural sequential modeling, etc. can be used to model and analyze observable EcoFin behaviors and events, their evolution, consequence, and impact. For example, techniques such as multi-task learning can be combined with text and document analysis to detect, summarize and classify financial events of different types from financial documents. Techniques such as Memoryless point process, Hawkes processes and Markov processes can be used to model sequential trading behaviors (actions such as buys and sells) and the transition and dependency between behaviors. In addition, AIDS techniques can be combined with behavioral finance and behavioral economics to explore and model the driving psychological factors for less rational to irrational human behaviors occurring in EcoFin markets and systems [32, 50, 71, 97, 114].

## 5.4   Multisource Economic-Financial Data Analysis

The comprehensive understanding of many EcoFin business problems including financial MTS analysis, cross-market analysis and enterprise-wide analysis has to involve multiple sources of data. Here, *multisource* broadly refers to data consisting of more than one type, including mixed types (e.g., numerical and textual), multiple modalities (e.g., trading transactions and news feeds), and multiple channels (multiple markets, products or services). Regarding the *EcoFin data structure* types, EcoFin data typically includes *transactional data* such as sales, marketing, trading, operational and service transactions; *textual data* such as long and short EcoFin text, e.g., EcoFin news, announcements, filings and reports; *visual data* such as images and diagrams embedded in financial news and documents; and *tabular data* such as statistical tables in financial reports and bank statements. From the *EcoFin product and system* perspectives, the analysis of an EcoFin problem may have to involve other EcoFin factors, indicators, products, services, markets and data to conduct joint analyses. From the *EcoFin data format* perspective, an EcoFin analysis may involve temporal and sequential data, longitudinal data, spatial and location-sensitive data, the tabular and matrix types of data, numerical data, categorical data, text strings, vector data, and the key-value type of data. In addition, both *macro-level* (e.g., macroeconomic data) and *micro-level* EcoFin data, and *local* (about a firm), *regional* (about a country or region) and *global* data may be involved in one task. Accordingly, many hybrid, ensemble and integrative analysis and learning tasks can be conducted on such mixed and multisource data for individual or multiple tasks and business objectives [109].

- Modeling the relevancy, redundancy and couplings (including correlations, dependency and causality) between financial indicators and features in multisource financial data [10, 12, 40].
- Integrating fundamental analysis and technical analysis by involving data about a stock or derivative's fundamentals and low-level technical indicators to predict its price, trend, movement and portfolio management [101, 144].
- Mixing structural and tabular (trading transactions), textual (news, announcements, filing and reports) and visual (images) features from the relevant multiple modalities and sources, jointly learning them for a representation of a financial asset, and then predicting the price, trend, movement or outlier of the asset; identifying top-ranked topics and stories about a stock corresponding to a given time series and interval/period [118, 120].
- Connecting an equity to its corresponding derivatives in both underlying and derivative markets and analyzing the bidirectional influence between the equity and its derived counterparties and context by modeling the cross-market connectivity and interactions [41, 184, 205, 206].
- Analyzing the couplings between a stock indicator's time series and its related event time, event types, event categories and stock movement-sensitive topics, concepts and dictionaries in financial news selected for specific security, sector, trend-sensitive events, topic or story of interest; predicting the impact of events and sentiment-sensitive statements in the news on the security or other assets such as foreign currency exchange (forex); predicting the price movement and selecting portfolios over specific or all time periods, time horizons, hierarchical time intervals, regions, sectors, and firms [127, 166].

- Transforming multi-factors in different sources of financial data and external data such as weather conditions, accidents (e.g., bushfires, storms and earthquakes) and healthcare data into vector, matrix, graph and other form-based representations for further learning tasks [118, 166, 191].
- Given financial review reports about a listed company, extracting and analyzing the accounting data, tabular data, compliance and regulation-sensitive features and events; checking the statement in the report against compliance rules; extracting suspicious, inconsistent, extra/emergent and counter-institutional statements, concepts, terms, and values [213].
- Given the financial whitepapers and reports about initial bitcoin and cryptocurrency offerings projects, extracting textual, numerical and accounting features about the offerings for the evaluation of investment opportunities and pricing; and combining report meta-data and layout features to predict financing amount and strategies [195, 210, 214].

In conducting the above tasks, many AIDS-related techniques can be involved to understand and analyze multisource financial data and address the above case studies. Several main approaches in analyzing multisource financial data are as follows. One is to convert heterogeneous financial inputs into one common embedding or representation, and then build learning models on this transformation for specific learning tasks, e.g., by deep GNNs [191]. Another is to analyze the heterogeneous inputs by respectively advantaged learning techniques and feed one result to another model for the overall learning tasks [103]. The third one is to build a learning model on the learning target while involving complementary data to tune the model or incorporate prior knowledge [118, 166]. Further, multivariate analysis is also a typical method to jointly analyze multiple variables from different financial time series and model their high-dimensional dependencies e.g. by copula functions [17, 198, 206]. Lastly, a base learner could be built for each source of data for the local learning task, and then the learned local results could be combined into a global result through an ensemble or a further transformation can be made before a classifier is connected to achieve the final learning target. Examples include integrating the autoregression of temporal data and text/document analysis techniques for financial text; building unified representations of mixed multisource input after transforming and concatenating the heterogeneous input data sources to vector or matrix-based representations for further transformation and learning tasks; learning the mathematical and statistical relations on financial text by techniques such as Bayesian nonparametric learning and then feeding the learned relations to time series forecasting [33, 135, 165]. Typical embedding and representation techniques include those for word and text representations, neural language models, time series regression, sequential representation, and network and graph-based representations by classic network and graph theories, NN, and neural graphical networks [18, 49, 54, 57, 96, 107, 110, 191, 196, 218]. In addition, the significance of different sources of data in contributing to a learning task can be modeled in terms of various types of attentions, such as multi-head attentions or hierarchical attentions [46].

## 5.5  Deep Financial Modeling

*Deep financial modeling* refers to the applications of deep learning methods in finance for the deep representations, abstraction and learning of EcoFin features, structures, distributions, relations, dynamics and effects. There could be many ways to categorize the relevant research. Here, we highlight three main directions: deep financial representation and prediction, deep cross-market, sector and factor modeling, and deep distributed financial modeling.

*Deep financial representation and prediction.* This has been the most intensively explored area which typically applies a basic DNN to a financial setting by utilizing the representation, abstraction and prediction power of DNNs. Four types of research are conducted on straightforward applications of deep learning. The first focuses on the *financial embedding, representation and modeling* of markets, financials and assets (e.g., securities, bonds and options), price formation mechanisms, volatility movement, supply-demand relations, market formation, including characterizing, extracting and selecting their representative features, commonalities, patterns, structures and relations. Their representation results from applying DNN such as RNN or combining DNN with classic stock representations such as linear representation or candlestick charts are applied to forecasting tasks such as stock movement and price prediction [49, 100, 107, 168]. The second is on *financial time series forecasting and deep sequential modeling*, which forms the main focus in various communities. Typical problems include price movement forecasting, volatility and return forecasting, market movement trend forecasting, portfolio construction, investment selection for stock markets (e.g., limit order book and daily data) and across sectors (sections) including indices, exchange rates, commodities, utilities and energy markets. Examples are to apply DNNs such as RNN, LSTM, GAN, CNN and regression neural networks or combine some of them with classic models such as time series regression, probabilistic modeling or classic learners to model financial time series for market movement trend and price forecasting, change detection or chart-based trend presentation [75, 137, 160, 161, 200, 217]. The third is to *develop trading strategies and signals* such as representing investor's trading behaviors, optimizing trading strategies, discovering trading signals and patterns and hedging opportunities, forecasting portfolios and constructing peer-to-peer lending and loan options, and enhancing trading strategies such as pairs trading. These can be achieved by methods such as applying DNNs like GANs and DRL like DQN together with classic methods such as time series analysis, text (announcements, social media or financial news) analysis, and evolutionary computing like genetic algorithms [57, 93, 110, 116, 117, 202]. The last relates to other applications including *deep risk estimation and management* e.g. for home loan and credit scoring, financial event detection, financial fraud detection, financial crisis analysis, and abnormal behavior analysis. Examples are applying methods like DNNs combined with classic financial models for value-at-risk, risk scoring, anomaly detection and change detection [39, 41, 88].

Well-established DNNs are applied to the above scenarios and tasks, such as restricted Boltzmann machines, deep convolutional networks, RNNs, autoencoders, GANs, DRL, and neural language models such as BIRT and Transformer, which model feature and temporal correlations and the sequential developments of financial inputs in a supervised or unsupervised mode [93, 104]. In applying DNNs, typical mechanisms including attentive networks, context embedding and dropouts are applied, in addition to other transformations such as scaling, filtering and normalization to neutralize and adapt to heterogeneous inputs [143]. In addition, DNNs are often combined with classic stock representation and modeling methods such as time series analysis, regression, candlestick charts and SVM etc. to jointly represent and model financial markets. The data involved may consist of single to multiple sources, e.g., stock tick or daily data, equity and derivative market data, financial reports and statements, financial news, and social media text. When multi-source or multi-modal data is involved, they are typically heterogeneously embedded (e.g., word embedding for news and financial reports, one-hot embedding for categorical features, or correlation matrix for multivariate numerical features) into DNNs for further homogeneous abstractions and transformations by the networks or are separately modeled by different networks (e.g., CNN or RNN for time series and Transformer for financial reports).

*Deep cross-market, sector and factor modeling.* Most EcoFin problems naturally involve various aspects of finance, economy and society and can only be understood well by jointly modeling the relevant markets, sectors and aspects of factors. Examples include involving different but relevant equity and derivative markets, exchange rate markets, commodity markets, digital currencies, lending, asset and wealth (e.g., gold and properties), utilities, energy markets, and other relevant factors (e.g., GDP). Deep cross-market, sector and factor modeling involves various research topics. The first is on the *deep correlation and dependence modeling* across markets e.g. stocks, indices, bonds, options, futures and commodities [172, 182, 194], across sectors e.g. energy, environment and high-tech securities [133, 218], and across macro-, meso- and micro-level factors e.g. GDPs, petrol prices, gold prices, digital currencies, and stocks [191]. The second is on *financial forecasting and prediction across markets/sectors/factors*, which predicts the trend or movement of one to multiple targets by involving other correlated or dependent ones. Examples are forecasting stock return by involving market index, GDP growth and job unemployment; recommending portfolio investment consisting of assets from multiple markets; and predicting financial crisis contagion across regions and sectors [4, 40, 41, 191].

The direct applications of DNNs consist of CNNs to model multi-market/sector/factor correlations and dependencies by embedding each factor as an input, RNNs like LSTM and bidirectional RNN to sequentially model multi-market/sector/factor time series, deep transfer learning to model cross-sectional investment strategies [4, 133], and integrating CNN (or RNN) with autoencoders, attention networks, memory networks and other mechanisms to jointly model sequential dynamics and other aspects of multi-markets/sectors/factors. A recent interest is in integrating DNNs (such as CNN, RNN, LSTM, GAN and autoencoders) with classic financial modeling methods (e.g. copula-based high-dimensional dependence modeling and ARIMA/GARCH-based time-series regression) [172, 182, 194] for market movement forecasting, portfolio investment prediction, and risk analysis.

*Deep distributed financial modeling.* An increasing number of financial businesses are delivered through distributed diverse platforms or services or involve multi-market/domain/channel/sector applications and services distributed across different locations. Typical examples are Internet finance, mobile payments, digitized asset management, crowd sourcing, and crowd-funding campaigns. The data-driven modeling of such financial businesses requires distributed and multi-domain/channel/sector analytics and learning. In addition to the above techniques for deep cross-market, sector and factor modeling discussed above, *deep distributed financial modeling* emerges as an increasingly important area of AI in finance. Typical learning techniques include distributed machine learning, deep transfer learning, federated learning, and cloud analytics [185, 208, 221]. Examples are to apply distributed and federated learning for privacy-preserving and secure resource allocation, multi-party financial applications, asset management, fraudulent credit card detection, and fraud detection [25, 209, 219].

## 6  GAPS IN THE AI RESEARCH IN FINANCE

### 6.1  Difference between AI and Finance Research

While both the AI and finance communities contribute to the overwhelming research on AI in finance, they demonstrate strong disciplinary preferences and objectives, resulting in practice differences and research gaps.

On one hand, the economics and finance communities are increasingly interested in applying AI in finance and FinTech. Many academia, industry and government-led forums and events have been organized on AI in finance and related topics, particularly on topics and areas of FinTech, smart payments, blockchain, and Internet finance. In the research, there is a long history of applying mathematical and statistical modeling techniques in finance and economics. Complex system methods, data analytics, machine learning, computational intelligence methods, etc. have also been widely introduced into EcoFin systems [24]. Typical characteristics of financial practice can be summarized as follows. (1) The applications focus on simple AI techniques and methods in finance and economics for more powerful financial explanations and insights, often ignoring a deep knowledge and understanding of the underlying AI techniques and their innovations. (2) AI is used as a complementary tool to explain, simulate and understand EcoFin phenomena, problems and mechanisms that are primarily dependent on EcoFin theories and tools. (3) Limited innovations of economics/finance-driven AI are available or focused. (4) It would not be surprising to see a very simple or even superficial understanding and applications of AI concepts and methods. (5) Small data is often used. (6) Results are evaluated and explained for the EcoFin purpose while systemic AI evaluation is of less concern.

On the other hand, AI has been applied in finance and economics for lasting success with increasingly stronger interest in broader, newer and smarter FinTech. This trend has been paramount and influential in driving today's data economy and smart FinTech. The relevant research demonstrates the following typical practice. (1) Innovation is focused on novel AI methods or novel AI applications in finance. (2) Finance is an application domain in the research, often without a deep understanding of the underlying domain and a domain-friendly interpretation of results. (3) Limited innovations are focused on improving EcoFin theories. (4) None to limited involvement of EcoFin theories and tools, and the problems are abstracted and characterized to fit models or data-driven discovery. (5) Large data is often involved. (6) Results are evaluated and interpreted in terms of comprehensive AI evaluation measures with less demonstrations on financial impact.

The above discussion inspires new opportunities and a need to (1) deepen the interdisciplinary understanding and knowledge of both AI and finance for smart FinTech synthesis; (2) transform AI and finance and develop cross-disciplinary synthetic theories and tools for smart finance, economy and society; and (3) develop economic/financial problems-oriented AI theories and tools to specifically address economic/financial problems and their characteristics and complexities. Typical complexities include diversified, nonstationarity and stylish effects, high-frequency data, market emergence, dynamics, uncertainty, extreme events and exceptions (such as black swan and grey rhino events [69]), high-dimensional and multi-aspect macro/micro-economic factors.

Table 2 further compares the difference between the AI and finance communities in terms of problem, data, method, evaluation, and result. Section 6.2 further discusses the gaps in the existing AI techniques when they are applied in finance. Section 7 further discusses the open opportunities to address some of the gaps identified in this section.

Table 2. Comparison of Interdisciplinary Practice in AI and Finance.

| Comparison | AI communities | Finance communities |
|---|---|---|
| Problem | AI research issues and challenges with applications in finance or inspired by financial problems and complexities | Financial problems and challenges by applying AI techniques or inspired by AI |
| Rationale | Complying with AI culture and theories, and advancing AI theories and applications | Complying with financial culture and theories, and advancing financial theories and interpretation |
| Data | Large economic/financial data with unique or high data complexities | Small and simple economic/financial data with rich interpretability |
| Method | AI innovations with advanced design and implementation | Innovations in financial theories or improved problem understanding |
| Evaluation | AI baselines and evaluation methods and measures | Financial baselines and evaluation methods and measures |
| Result | Achieving significant statistical and technical performance or effect improvement in financial applications | Better results or new perspectives for financial explanation |

## 6.2   Pros and Cons of AI Research in Finance

In this section, we comment on the pros and cons of the typical AIDS techniques and methods discussed in Sections 4 and 5 which are widely applicable to EcoFin problems and systems in terms of six major families of AI techniques in finance [33]: (1) mathematical and statistical modeling, (2) complex system methods, (3) classic analysis and learning methods, (4) computational intelligence methods, (5) modern AIDS methods, and (6) hybrid AIDS methods, as discussed in Section 4.

First, *mathematical and statistical modeling* forms the AI foundation to characterize, formulate, model, analyze and optimize EcoFin systems and their working mechanisms, problems and solutions [3, 5, 23, 55, 77, 108, 126, 153, 160, 167, 175]. Table 3 summarizes the typical AI-oriented mathematical and statistical techniques and methods and their advantages and disadvantages in financial applications.

Table 3.  Mathematical and Statistical Techniques and their Pros and Cons in Financial Applications.

| Techniques | Methods | Pros in finance | Cons in finance |
|---|---|---|---|
| Numerical methods | Linear and nonlinear equations, least squares methods, finite difference methods, dependence modeling, Monte-Carlo simulation, etc. | Model-driven, hypothesis testing, and forecasting; mathematically modeling determinant financial processes, mechanisms and dynamics; analytic or approximate results and interpretation; etc. | Complex processes, mechanisms and dynamics; high-dimensional/order and low-quality (e.g., missing, incomplete, inconsistent) data; nonstationary, heterogeneous, dynamic, uncertain and large data; population size; etc. |
| Time-series and signal analysis | State space modeling, time-series analysis, spectral analysis, long-memory time-series analysis, nonstationary analysis, etc. | Modeling temporal processes, relations, dynamics and effects; trends, movements, changes and forecasting; multivariate relations and movements; etc. | Non-temporal, multiple and heterogeneous relations, processes and dynamics; mixed factors, data, relations and processes; poor data quality (e.g., noise) and stylist effects; structure and sample dynamics and nonstationarity; overfitting; population size; etc. |
| Statistical learning methods | Random walk models, factor models, stochastic volatility models, copula methods, nonparametric methods, etc. | Model-driven and hypothesis testing; sampling; latent variables, relations and models; dependency, uncertainty and randomness; probabilistic interpretability; etc. | Modeling other and mixed relations, processes and dynamics; mixed observable data; poor-quality data; large data and scalability; result actionability; etc. |
| Random methods | Random sampling, random walk models, random forest, stochastic theory, fuzzy set theory, quantum mechanics, etc. | Modeling random processes, relations and dynamics; randomness, uncertainty, fuzziness; fair and unbiased representativeness; etc. | Too small or large populations; complex (e.g., imbalanced, unequal) data; dynamic data; mixed data complexities; bias and error; etc. |

Second, the theories of *complex systems* have been intensively applied in classic AI-driven finance and economics. They are applied to understand, simulate and visualize the system complexities (e.g., self-organizing behavior and emergence of intelligence) and working mechanisms of complex EcoFin systems, processes, and problems [29, 53, 66, 81, 83, 113, 132, 173, 176, 188]. Table 8 summarizes the typical AI techniques and methods enabled by the theories of complex systems and their advantages and disadvantages when used in financial applications.

Table 4. Theories of Complex Systems and their Pros and Cons in Financial Applications.

| Techniques | Methods | Pros in finance | Cons in finance |
|---|---|---|---|
| Complexity science | Systems theory, complex adaptive systems, chaos theory, random fractal theory, etc. | Modeling system complexities; simulating complex financial processes, mechanisms, and characteristics; incorporating social science; trial andtest; etc. | Incomplete and limited understanding; mixing qualitative and quantitative factor modeling; evaluation and optimization; etc. |
| Game theory | Zero-sum game, differential game, combinatorial game, evolutionary game, Bayesian game, etc. | Hypothesis testing; rule-based design; scenario analysis; trail and test; involving reinforcement learning; etc. | Complex design and interactions; involving financial rules, theories and measures; complex financial characteristics; etc. |
| Agent-based modeling | Multiagent systems, belief-desire-intention model, reactive model, swarm intelligence, reinforcement learning, etc. | Rule-based modeling and controlled experiments; simulation; self-organizing mechanisms; mechanism testing; etc. | Complex financial scenarios; runtime behaviors and decision-making; dynamics; quantitative and formal modeling; etc. |
| Network science | Linkage analysis, graph methods, power law, small worlds, contagion theory, etc. | Modeling networking modes, interactions and processes; social influence; visualization; etc. | Complex interactions, relations and processes; heterogeneous and high-dimensional financial systems and mechanisms; multi-aspects, sources and modalities; etc. |

Third, *classic analytics and learning methods* form a large part of the classic AI and machine learning research. They have been widely applied in modeling, characterizing, analyzing, learning and discovering interesting knowledge, forecasting trends and movements, and detecting outliers in EcoFin data, behaviors, events, processes and systems [27, 36, 37, 42, 68, 74, 99, 216, 220]. Such learned knowledge is then used to optimize and manage EcoFin events, behaviors and systems. Table 5 summarizes the typical AI techniques and methods enabled by the theories of classic analytics and learning methods and their advantages and disadvantages when used in finance.

Further, *computational intelligence* techniques and methods inspired by nature and human systems and simulating natural and human intelligence are used to model, characterize and optimize EcoFin mechanisms, strategies and systems and to manage and optimize portfolio and market forecasting, etc. [43, 60, 80, 86, 105, 122, 123, 146, 147, 150]. Table 6 summarizes the typical techniques and methods falling in the computational intelligence area and their advantages and disadvantages when used in financial applications.

Table 5. Classic Analytics and Learning Methods and their Pros and Cons in Financial Applications.

| Techniques | Methods | Pros in finance | Cons in finance |
|---|---|---|---|
| Pattern mining methods | Frequent itemset mining, sequence analysis, combined pattern mining, high-utility pattern mining, tree pattern, network pattern, knot pattern, interactive pattern, etc. | Extracting and recognizing patternable structures, modes and relations; interpretability; actionable rules; etc. | Infrequent events and behaviors; high false positive; missing important financial events and activities; relations and interactions; etc. |
| Kernel learning methods | Vector space kernel, tree kernel, support vector machine, spectral kernel, Fisher kernel, nonlinear kernel, multi-kernel methods, etc. | Kernel tricks for complex problem modeling; scalable; metric properties; etc. | Kernel selection; calibration; training; interpretability; and actionability; etc. |
| Event and behavior analysis | Sequence analysis, Markov chain process, high-impact behavior, high-utility behavior, nonoccurring behavior analysis, etc. | Modeling financial events, processes, interactions, and activities; sequential modeling; event/behavior impact; simulation; visualization; actionable; etc. | Complex processes, interactions and behaviors; insufficient event/behavior modeling theories and tools; mixed and dynamic processes; uncertainty and change; predictive consequences; etc. |
| Document analysis and NLP | Language models, case-based reasoning, statistical language model, Bayesian model, latent Dirichlet allocation, Transformer, BERT, etc. | Linguistic and semantic representation and analysis; unstructured data; long and short text; subjective and objective factors; sentiment; opinion; etc. | Complex syntactic and semantic relations and structures; expression informality; financial knowledge embedding; multi-linguistic interactions and heterogeneities; scalability; etc. |
| Model-based methods | Probabilistic graphical model, Bayesian networks, expectation-maximization model, clustering, classification, deep neural models, etc. | Hypothesis testing; modeling design-based processes, behaviors and dynamics; paradigm and modeling tools-dependent; wide applicability; etc. | Runtime and dynamic modeling; handcrafting issues; model selection; under/over-fitting; financial semantics and knowledge; financial explanability; etc. |
| Social media analysis | Topic modeling, sentiment analysis, emotional analysis, influence analysis, linkage analysis, interaction learning, etc. | Modeling financial events, sentiments and opinions; social activities, interactions and influence; etc. | Misinformation; bias; mixed financial data and information; financial mechanisms, processes and systems; etc. |

Table 6. Computational Intelligence Methods and their Pros and Cons in Financial Applications.

| Techniques | Methods | Pros in finance | Cons in finance |
|---|---|---|---|
| Neural computing methods | Wavelet neural network, genetic neural network, recurrent neural network, deep neural network, etc. | Simulating human neural and cognitive systems and mechanisms; complex relations and structures; incomplete input; data-driven modeling; fault-tolerance; distributed/parallel processing; hierarchical processing; great learning performance; etc. | Limited to neural mechanisms; other complex processes and mechanisms; network structure selection; parameterization; overfitting; computation-intensive; interpretability; action-ability; etc. |
| Evolutionary computing methods | Ant algorithm, genetic programming, self-organizing map, artificial immune system, swarm intelligence, neural-genetic algorithm, etc. | Simulating evolutionary and genetic systems and mechanisms; financial simulation; optimization; exploratory results; etc. | Limited to evolutionary mechanisms; bias in local/global optimal and extremum; configuration and parameter tuning; etc. |
| Fuzzy set methods | Fuzzy set theory, fuzzy logic, fuzzy neural network, genetic fuzzy logic, etc. | Modeling uncertain, imprecise, fuzzy and contradictory inputs; contradictory objectives; linguistic support; rule/symbol-based interpretation; etc. | Modeling limit; fine tuning; approximation; data-driven modeling; complex financial mechanisms, relations and processes; etc. |

Table 7 lists several typical *modern AI techniques* and their representative methods. Such modern AI technologies play increasingly critical roles in deeply understanding, modeling, analyzing, predicting, optimizing and managing complex EcoFin problems and systems [8, 44, 47, 51, 54, 77, 79, 82, 99, 105, 108, 109, 129, 138, 141, 145, 159, 160, 164, 183, 192]. The table summarizes their representative techniques and methods and their advantages and disadvantages when used in financial applications.

Lastly, real-life EcoFin systems and problems become increasingly complicated, and their effective and efficient problem-solving often requires the hybridization of multiple complementary classic and modern AI techniques and methods [16, 29, 65, 71, 97, 103, 109, 148, 211]. Table 8 illustrates several *hybrid AI methods* and their advantages and disadvantages when used in financial applications.

The above discussion on the pros and cons of existing AI techniques in finance discloses the opportunities for disciplinary and technical development of future AI and finance and their synthesis. In the following section, we address some of the weak areas and novel opportunities for the smart future of AI in finance.

## 7 OPEN OPPORTUNITIES: SMART FUTURES

The open opportunities for AIDS in EcoFin are to empower smart EcoFin, which can be broadly categorized into three families highlighted below: (1) *AIDS-empowered EcoFin developments* where AIDS drives the smart futures of economics and finance; (2) *EcoFin-driven AIDS developments* where current and future demand and developments of economics and finance further inspire new AIDS techniques for smart EcoFin [156]; and (3) *beyond AIDS technologies* where other opportunities go beyond the current technical scopes and foci of AIDS in EcoFin. Below, we discuss these perspectives in terms of AIDS-driven strategic planning and developments for smart EcoFin, AIDS-enabled EcoFin Innovations, and opportunities beyond AIDS.

Table 7. Modern AI Techniques and their Pros and Cons in Financial Applications.

| Techniques | Methods | Pros in finance | Cons in finance |
|---|---|---|---|
| Representation learning | Probabilistic model, graph network, network embedding, tree model, neural embedding, etc. | Characterizing intrinsic characteristics and relations; discovering discriminative features and representations; various methods; enabling learning tasks; etc. | Representation constraints; curse of dimensionality, relation, interaction and heterogeneity; unsupervised learning; interpretability; etc. |
| Short and informal text analysis | Conceptualization, term/tag/phrase similarity learning, dependency parsing, word embedding, deep neural models, etc. | Analyzing informal communications and presentations; contextual and sequential presentations; linguistic specificity; etc. | Informality complexities; limited information; noise and misinformation; uncertainty and inconsistency; etc. |
| Optimization methods | Nonlinear, stochastic and dynamic programming, information theory, Bayesian optimization, etc. | Theoretical guarantee of correctness and generalization; hypothesis and setting specific; optimal results; etc. | Constraint; approximation; local/global optimum; computational complexity; applicability in complex financial systems; etc. |
| Reinforcement learning | Bellman Equation, actor-critic model, Markov dynamic progress, deep Q-network, adversarial reinforcement learning, etc. | Learning from mistake and correcting error until perfection; scenario/context-based policy and action optimization; modeling interactions; balancing exploration and exploitation; etc. | Markovian assumption; data and computation-hungry; scenario and policy specific; long-term dilemma; curse of dimensionality and samples; etc. |
| Deep learning methods | Convolutional neural network, attention network, generative adversarial network, autoencoder, deep Bayesian network, etc. | Deep abstraction and representation; powerful learning architectures; complex hidden relations and interactions; domain knowledge and supervised information; impressive learning performance; etc. | Data and computation-intensive; observable and explicit financial complexities; robustness; interpretability; actionability; etc. |
| Deep financial modeling | Deep neural networks, deep reinforcement learning, deep language learning, hybrid DNN with ARIMA, GARCH and copula etc. classic financial models, federated learning, etc. | Deep representation and learning of correlations, dependencies and temporal developments between financial factors and across financial markets and sectors | Vulnerability of modeling small financial data with asymmetric, non-normal, nonstationary and abruptly evolving features, heterogeneous distributions, and hierarchical couplings within and between implicit and explicit factors, etc. |

Table 8. Hybrid AI Techniques and their Pros and Cons in Financial Applications.

| Techniques | Methods | Pros in finance | Cons in finance |
|---|---|---|---|
| Parallel ensemble | Evolutionary neural models, ensemble learning, deep Bayesian model, copula graph neural network, combining complexity science and game theory, etc. | Jointly modeling two to more financial aspects; integrating model capabilities; wide applicability in finance; better performance; etc. | Model selection and integration; compatibility; customized modeling process and structure; problem/task-specific; complexity; etc. |
| Sequential and hierarchical hybridization | Time-series analysis plus classification, macro/micro-economic dependency modeling, deep sequential modeling-based event detection, etc. | Jointly modeling two to more financial aspects; enhanced modeling capabilities; multistage modeling; wide applicability in finance; better performance; etc. | Model selection and integration; compatibility; customized modeling process and structure; problem/task-specific; complexity; etc. |
| Cross-disciplinary hybridization | Deep multi-time series analysis, copula deep models, autoregression deep model, behavioral economics and finance, etc. | Jointly modeling two to more financial aspects; enhanced modeling capabilities; wide applicability in finance; better performance; etc. | Method selection and integration; compatibility; customized modeling process and structure; problem/task-specific; complexity; etc. |
| Behavioral economics and finance | Prospect theory, nudge theory, natural experiment, experimental economics, behavior informatics, intention learning, next-best action modeling, etc. | Modeling psychological drivers and insights of EcoFin behaviors; financial experiments; subjective and social factors; interactions, behaviors and actions; etc. | Generalized theories and tools; qualitative-to-quantitative modeling; objective and subjective modeling; scenarios and samples specific; predictive power; etc. |

## 7.1 AI-driven Strategic Planning and Development

*Strategic planning and development* determine the short to long-term vision, missions, strategic plans, actions and strategies of an organization, and optimize the systematic and harmonious identification, alignment, communication, collaboration and cooperation between some fundamental organizational elements (e.g., strategy, structure, system, shared value, skills and competencies, leadership style, and staff and their capabilities). Therefore, the quality and implementation of strategic planning and development critically affect the position, operations, performance, competitiveness and future of the organization. In EcoFin, strategic planning and development play an even more critical role than the other tasks and agenda. Below, we illustrate how AIDS could significantly lift and benefit the strategic planning and development for EcoFin organizations, businesses and the economy of a country or region in terms of common issues in strategic planning and implementation, specific aspects of a financial organization's strategic planning, and opportunities in managing the changing nature of EcoFin systems. These may involve AIDS and other related techniques including long and short text analysis, information extraction and retrieval, event analysis, mixed data analytics, and sequential modeling.

- Developing the interpretability of strategic visions, missions, initiatives and implementation strategies and actions of a company; providing evidence, estimation, forecasting and simulation to support, explain and guide the strategic planning and their implementations.
- Evaluating the soundness, applicability, feasibility and performance of an organizational strategic plan by benchmarking against its actual annual historical plans, developments and performance and against the practices, implementations and performance of the same-sector and same-sized companies; and optimizing the strategic planning and development.
- Reviewing the performance gaps between the proposed long-term organizational vision, short-to-middle term missions and the actual annual objectives, practice and outcomes by techniques such as text analysis and key indicator comparison of an enterprise's strategic plan and its financial statements and annual reports; and adjusting and optimizing the organization's strategic planning and development.
- Reviewing the alignment and gaps between a corporate strategic plan and departmental plans, objectives, targets, means and implementation methods, agenda, processes, and periodical performance and outcomes; reviewing the pros and cons of enterprise plan against departmental plans; reviewing the gaps and unalignments between upper managerial vision and plans and middle-to-lower unit's day-to-day management and operational activities; and optimizing the alignment and execution of both enterprise and departmental strategic plans and hierarchical performance expectations of managements and departments.
- Reviewing the leadership, performance, strategy and implementation inconsistencies and gaps between engaged and disengaged departments and employees, between high-performing and low-performing units and individuals, between strong and weak leaderships, and between business partners; and optimizing the whole-of-enterprise quality and performance of strategic planning, cooperation, and implementation.
- Reviewing the harmony between an organization's strategic plans and their implementation goals, measures (KPIs), initiatives, processes, finance, customers and capacity and growth; and optimizing their connectivity, alignment and executional performance.
- Reviewing the effectiveness of initiatives that are created for achieving the objectives, and adjusting and optimizing the initiatives.

Specifically, the strategic planning and development for an EcoFin business or organization may require designing, implementing, evaluating and optimizing the strategic plans and their implementation strategies and methods, measures, programs and projects, products and markets, financials, staffing, capacity and capabilities, operations and review. Below, we list a few examples.

- Analyzing, adjusting and optimizing the strategic directions and objectives suitable for a financial market, product or service; and designing the aligned implementation strategies, methods and evaluation measures.
- Interpreting strategic decisions and measures, e.g., key company merger, market movement, credit rating and corporate risk, in terms of supporting evidence, statistics and simulation; and estimating the impact on affected business and its performance and results.
- Identifying and discovering the strategic markets, channels, major clients and users; and designing and deploying the suitable financial products, services, and marketing campaigns.
- Discovering strategic customers, particularly for new and future financial products and services; and estimating their sensitivity to product design, performance, pricing, and service level and quality.
- Estimating the lifespan of an EcoFin system, product or service, particularly for new and future products and services; simulating the product life cycle; and creating suitable staffing, marketing, financial and supporting strategies, methods and facilities.

- Estimating and simulating the new intent of consuming existing EcoFin products and services and changing consumption interests, preferences, habits, spending, and expectations on EcoFin products and services for high-value and high-potential customers; and recommending strategic plans and developments to cater for early churning, attribution, service upgrading and advanced customer care.
- Analyzing and detecting sensitive and high-value multi-policy and multi-product clients of EcoFin products and services; recommending marketing campaigns and customer care activities for new and existing high-value clients.
- Conducting market analysis and competition analysis of existing and new EcoFin products and services in a market in terms of their SWOT (strengths, weakness, opportunities and threats) against their own products and services; and recommending optimization and mitigation strategies and activities to improve market share, competitivity, quality of service and the performance of their own products and services.
- Reviewing the financial product roadmap from design to production, sales and marketing in terms of product performance, market compactivity, market share dynamics, customer feedback, the market lifecycle of the product, and contextual factors in the market.
- Evaluating the consumer sensitivity to EcoFin products and services in terms of their branding, design, innovation, pricing, layout, functionality, usability, market and channels, marketing, sales and customer relationship management in comparison with non-financial products and services and alternative products and services; and optimizing the positioning, design, packaging, marketing and services of competent EcoFin products and services.
- Categorizing the determinants of investor emotion and sentiment in equity, derivative, commodity, foreign exchange, property, insurance and lending markets; discovering the sentimental difference of the same customers on different EcoFin products and services.
- Analyzing the influence of confounders, group users and contextual users on the sales and marketing of EcoFin products and services; and recommending EcoFin products and services to those influence-sensitive customers and organizations.
- Analyzing the independence level of customers and organizations on EcoFin products and services; recommending EcoFin products and services to those self-motivated and independent consumers and organizations; and suggesting independent customer-oriented personalized product design, marketing, and services.

In EcoFin strategic planning, *understanding and managing changes in economic-financial systems* [5, 94, 111] is more common and challenging than in other sectors and businesses. EcoFin systems, products and services are more connected to and deeply influenced by the rest of world, including market dynamics and sentiment, politics, globalization and exceptional events, than many conventional businesses such as manufacturing and retail services. Both global and local changes, demand and supply changes, subjective and objective changes, and planned and unexpected changes could significantly affect EcoFin products and services.

- Understanding the changes of company culture, governance and management; quantifying their impact on corporate strategic planning and implementations, e.g., by analyzing the financial statements, annual reports and company statements to identify the changes in terms of word, phrase or jargon usage variations in company strategic planning reports, and financial texts.
- Modeling the influence of the unexpected movement of key management, staff and specialists, the adjustments of management strategies, processes and systems, and the change of production plans, procedures and business partnership; and making recommendations for strategies and treatments to manage the downside impact and enhancing the positive effect.

- Modeling changes in terms of diachrony (time evolution) and synchrony (variation across sources and authors), disparity and evolution of visions, opinions and arguments of financial actors and participants in annual financial reports and bank statements.
- Detecting the exceptional number of external accidents and events (significant climate accidents, geographical events, epidemic accidents, political events, and terrorist activities) and their influence on internal planning, objectives, operations, and implementations.
- Analyzing and detecting the change in the quality of products and services and the change in customer feedback and sentiment; predicting the churning of customers across companies and competitors in the sector; and recommending intervention and customer care strategies and timing.
- Detecting the change of intent, emotion, sentiment and preferences on existing financial products and services with a strong possibility of seeking new and alternative products and services; and predicting alternatives of interest.
- Analyzing and predicting the changes of customer circumstances, financial conditions, consumption demand and habits, and new interest and needs (e.g., becoming richer with substantially more income and earnings, demanding more advanced and luxurious products and services, varying interest from low-level services to high-level services, or becoming inactive and bored with some products and services); recommending personalized high-end financial products and services and wealth management strategies to high-end customers.
- Analyzing and predicting a change in customer income and revenue, circumstance, and the demand on credit, loan and lending products and services; predicting their refinancing need and asset/wealth growth; and recommending intervention and treatments.

## 7.2 AI-enabled Economic-Financial Innovations

The advancement of new-generation AIDS, machine learning and deep learning drives the paradigm shift from conventional economy and finance (which are driven by social science and economic and finance theories and methods) to the new era of economy and finance (that widely involves deep data analysis, data-driven evidence discovery, and combines data-driven discovery and machine learning theories with economic and finance theories) [14, 136, 190]. Significant innovation opportunities spread from upgrading and transforming classic EcoFin systems, theories, products and services and addressing significant real-life opportunities and complexities in economy, finance, society and environment to creating new EcoFin systems, products and services.

In the scenarios of multi-factor, multi-instrumental, multivariate, multi-source, multi-view, multimodal, multi-distributional, high-dimensional cross-market and cross-sectional EcoFin data and businesses, the real-life and contextual analytical and learning paradigms for smart FinTech shift from the entire-object and cohort analysis typically based on the IID assumption to addressing the above real-life EcoFin data complexities. For example, various couplings, interactions, communications, complementation, conflict and constraints exist within each aspect of EcoFin factors, between aspects of factors and with the target problems. Accordingly, a more feasible and effective methodology to address such real-life complexities in the new economy and finance is to combine the bottom-up thinking approaches (individualism) with top-down thinking and approaches (holism) to create systematic thinking and modeling (systematism) [29]. Systematism seeks a balance between individually personalized but contextually generalized results, between point-wise stationary but scenario-wise nonstationary (including diversity, dynamics, significant changes and variances), between aspect-specific privacy (e.g., individual characteristics, structure and influence) and systemic openness/publicity (e.g., communications and networking between individuals), between singular (only consider single aspect) and multiplex (heterogeneous, compound and hierarchical) mode, between decoupled (factors and aspects are independent or disentangled) and coupled (interactive, interconnected, entangled, and mutually influential), and between full observability (visible observations) and partial-to-full invisibility (known and unknown hiddenness, e.g., implicit or hidden factors, aspects, relations, structures and distributions, and unrecognizable by specific methods). Below, several examples are listed which address these EcoFin demands and challenges.

- Modeling, analyzing and predicting EcoFin problems by considering the real-life data characteristics and complexities, including data scale, coupling relationships, dynamics and changes (drifts and bursts), large-scale, high-dimensionality, higher-order distributions, multi-distributions, nonstationarity, stylist, inconsistency, uncertainty and stochastic features of data and business, as well as various other data quality issues.
- Modeling, analyzing and predicting EcoFin problems by considering real-life market circumstances, class imbalance, largely unlabeled cases, novel cases, self-motivated and evolving behaviors of market participants and regulators, market contexts, and contextual factors.
- Considering modeling capacity and capability limitations in handling complex EcoFin data and market characteristics, including overfitting or underfitting, a weak generalization capability, soundness, adaptive capability, and active learning capability.
- Evaluating the modeling performance of conventional EcoFin models such as ABM, classic reinforcement learning models such as game theory, and optimization such as bandit optimization in handling significant real-life and the new challenges of increasingly complex and novel EcoFin systems, products and activities.
- Improving conventional multiagent reinforcement learning such as actor-critic reinforcement learning for portfolio management to address local minima, multi-objective optimization and asset couplings by learning policies from data to guide actors for optimal actions and learning the market impact from data to improve reward; analyzing historical trading behaviors and patterns to supervise the agent action-taking and agent state-updating; and developing new optimization methods to upgrade policy gradient and evolutionary algorithm to fit the complexities of real-life EcoFin optimization.
- Reviewing the roles, performance, experiments and evaluation of AIDS-enabled financial engineering, their reproducibility, trustfulness and interpretation of learning results, the meaningful evaluation measures, the deliverable format and actionability of AIDS-enable results for actionable implementation and interpretation in production and decision-making; generating machine learning and data science models and systems that are accurate (can predict well), transparent (can be explained and implemented in human understandable and business actionable rules) and ethical (compliant and conforming to financial laws, regulations and privacy protection practice) in relation to EcoFin theories and businesses.

In addition, AIDS and related techniques such as deep learning and data science will substantially transform and upgrade the current practice of smart economy and smart finance [1, 13, 14] in terms of addressing many specific innovation projects and tasks, such as the following examples.

- Building universal representations of EcoFin systems, products and services, e.g., stock representation, index representation, currency representation and the representation of a company or a country's economy by involving various relevant EcoFin factors, indicators and contextual measures.
- Modeling individual to compound EcoFin targets (e.g., loan default, credit score, return over investment, etc.) by considering the context, which may involve other loosely or tightly relevant and coupled products and services, underlying policy and system constraints, market circumstances and trends, macro-level EcoFin factors and events, and dynamics and changes of contextual factors.

- Catering for distributed learning with data and models distributively owned and updated by data and model owners; enabling blockchain-based distributed learning to support decentralized, parallel and secure data and model sharing; building the proof-of-work consensus on models and protecting data privacy and owner anonymity; as well as coping with inconsistent dynamics of data, model, users and their changes.
- Automating investment by building automated trading, financing and advising systems and services that can collect the relevant data online and from third-parties, feed private data about a company or investor, analyze the profile, risk, performance and growth, and predict and recommend advice for investment strategies, portfolio optimization, and risk mitigation.
- Constructing, selecting and discovering more discriminative features from EcoFin data and business, e.g., suitable technical indicators, time series, fundamental indicators and news as well as their timespans and horizons, for developing more effective and actionable EcoFin analytics and learning, e.g., for stock movement prediction and anomaly detection.
- Developing automated trading systems to conduct both EcoFin theory-based fundamental analysis and technical analysis; incorporate machine learning methods to analyze and optimize a portfolio of assets by considering fundamental and technical indicators; generate automated pricing, signal generation, order placement, portfolio selection, strategy recommendation; and market impact and risk analysis.

### 7.3 Beyond AI and Finance

This section discusses open issues and opportunities beyond technologies for next-generation AIDS-enhanced EcoFin. We highlight the following six aspects: strategic and innovative thinking, broader EcoFin areas, neuropsychological foundation, social and ethical issues, geopolitical and geocultural influence, and the economics of AIDS-enabled economy and finance for smart EcoFin.

*AIDS-empowered strategic and innovative thinking for EcoFin.* Following the discussion in Section 7, a main goal of next-generation EcoFin is to make EcoFin 'intelligent', i.e. smart EcoFin. Smart EcoFin is embodied in (1) translating centralized, state-owned and institutional EcoFin infrastructure, systems, products and services to more distributed, personalized, small businesses and human-centric, shareable and open economy and finance; (2) translating authoritative, hypothetical, predefined and heuristic EcoFin theories and tools to evidence-based, verifiable, adaptive, risk-tolerant and actionable planning, operations, production and regulation; (3) translating trial-and-test and design-time-based developments to be proactive and run-time-based; and (4) translating fundamental demand-driven EcoFin innovations and supply to personal demand-driven innovations and supply. All such translations need to be built on the new-generation AIDS, and the corresponding EcoFin innovations will drive the AIDS innovations to address needs and requirements.

*AIDS for broader economic-financial areas.* Though this paper focuses on the mainstream EcoFin businesses and areas, AIDS have also widely contributed to many other areas, e.g., social good, job growth, productivity enhancement, well-being grow, risk and security mitigation, and society development. New EcoFin innovations will open tremendous new and unprecedented opportunities in these areas, enabling anytime, anywhere, anybody and any-form EcoFin demand. A big question to ask is what in economics and finance can be translated and what can be renovated by new-generation AIDS, e.g., the creation of EcoFin robotics in investment-advising, marketing, finance, daily business operations, risk management and compliance.

*Neuropsychological foundation for AIDS and EcoFin.* Both EcoFin and AIDS advancements and innovations require a deep understanding of how humans, nature and the universe work and evolve. As artificial systems, smart EcoFin futures have to explore neuroscience and psychology to understand, represent and formulate human cognitive, emotional, subjective and situated perception, thinking, learning, problem-solving and decision-making mechanisms. Both EcoFin and AIDS need to build their neuropsychological foundations, e.g., neuroscience-enabled AIDS, leading to new characterization, representation and learning of how the nature and universe develop and evolve. The findings will advance EcoFin accordingly, and new forms of EcoFin systems and services will then emerge and boom.

*Social and ethical issues in AIDS-driven EcoFin.* Smart EcoFin not only brings about socioeconomic benefits to the society advancement but also incurs increasing social and ethical issues. Examples are competition fairness and transparency, price inequality, resource and human power plunder and control; the robustness, sustainability, security and privacy of AIDS for economy and finance; the transparency, explainability and trustworthiness of AIDS for the economy and finance; the quality and performance of AIDS-enabled economy and finance; and enhancing development inequality and unsustainability. AIDS-empowered EcoFin also needs to develop computational regulation and compliance and settle regulatory rules for mitigating negative impact of AIDS applications, e.g., the misuse and abuse of intelligence, robotics and automation in EcoFin systems and activities, e.g., manipulating currency rate, intellectual property and intangible assets. All of these also depend on the development of novel AIDS.

*Geopolitical and geocultural influence.* Any EcoFin systems and activities are influenced by geopolitics and geoculture. Smart EcoFin aims for not only universal EcoFin intelligence and advancements but also geopolitical and geocultural advantages and personalization. AIDS can enable the understanding of geopolitical and geocultural characteristics, complexities, advantages, and the development of geopolitically and geoculturally smart EcoFin. Examples are understanding unique political systems, culture, language, and geographical differences and embed such uniqueness and differences into smart EcoFin. This will also drive AIDS innovations.

*The economics of AIDS for economy and finance.* This is to study the economic roles of AIDS as a general-purpose technology for EcoFin development and innovations; the economic influence of AIDS on businesses, labor, productivity, compatibility, jobs, growth, futures and EcoFin research; the socioeconomic impact of AIDS on the economy, finance, society and wellbeing [91]; and the mitigation and regulation of AIDS-empowered EcoFin systems, activities, change and consequences, etc. These studies will require the interdisciplinary innovations of both AIDS and EcoFin.

## 8   CONCLUSIONS

AI in finance has been a continued substantial research direction over decades with increasing cross-disciplinary interactions and fusion between AI, data science, machine learning, finance and economics. This trend has been further enhanced in recent years with the fast development of new-generation AI and data science and their applications to broad-based financial applications. This review presents a comprehensive and dense overview of the advantages and weaknesses of classic and modern AIDS techniques in finance. In particular, we review and comment on the data-driven methods in financial applications. The review further triggers discussion on the open issues and future opportunities of new-generation AIDS in finance and their synergy. This review significantly leverages many related reviews where only specific AI methods or financial problems are the focus. It also complements our other systematic review on the widespread financial applications benefiting from classic and modern AI research.

## ACKNOWLEDGMENTS

This work is partially sponsored by the Australian Research Council Discovery grant DP190101079 and the ARC Future Fellowship grant FT190100734.

†

---

†Note, due to space limit, only most relevant reviews, surveys and references are listed here as there are simply too many papers related to the very broad topics in this paper. More on AI in FinTech is at https://datasciences.org/fintech/.